\documentclass[prb,aps,twocolumn,showpacs,amsmath,amssymb,floatfix,superscriptaddress,citeautoscript]{revtex4-1} 
\usepackage{graphicx}
\usepackage{amsmath}
\usepackage{amssymb}
\usepackage{times}

\newcommand{\Tk}{T_{\rm K}}

\usepackage{color}

\begin{document}

\title{Mesoscopic Anderson Box: Connecting Weak to Strong Coupling}

\author{Dong E. Liu}
\email{dl35@phy.duke.edu}
\affiliation{Department of Physics, Duke University, Box 90305,
  Durham, North Carolina 27708-0305, USA} 
\author{S\'{e}bastien Burdin}
\affiliation{Univ Bordeaux, LOMA, UMR 5798, F-33400 Talence, France, EU}
\affiliation{CNRS, LOMA, UMR 5798, F-33400 Talence, France, EU}
\author{Harold U. Baranger}
\affiliation{Department of Physics, Duke University, Box 90305,
  Durham, North Carolina 27708-0305, USA} 
\author{Denis Ullmo}
\email{denis.ullmo@u-psud.fr}
\affiliation{Univ Paris-Sud, LPTMS, UMR8626, 91405 Orsay, France, EU}
\affiliation{CNRS, 91405 Orsay, France, EU}
\begin{abstract}
 We study the Anderson impurity problem in a mesoscopic setting,
 namely, the ``Anderson box'' in which the impurity is coupled to
 finite reservoir having a discrete spectrum and large
 sample-to-sample mesoscopic fluctuations. Note that both the weakly
 coupled and strong coupling Anderson impurity problems are
 characterized by a Fermi-liquid theory with weakly interacting
 quasiparticles. We study how the statistical fluctuations in these
 two problems are connected, using random matrix theory and the slave
 boson mean field approximation (SBMFA). First, for a resonant level
 model such as results from the SBMFA, we find the joint distribution
 of energy levels with and without the resonant level present. Second,
 if only energy levels within the Kondo resonance are considered, the
 distributions of perturbed levels collapse to universal forms for
 both orthogonal and unitary ensembles for all values of the
 coupling. These universal curves are described well by a simple
 Wigner-surmise-type toy model. Third, we study the fluctuations of
 the mean field parameters in the SBMFA, finding that they are
 small. Finally, the change in the intensity of an eigenfunction at an
 arbitrary point is studied, such as is relevant in conductance
 measurements. We find that the introduction of the strongly-coupled
 impurity considerably changes the wave function but that a
 substantial correlation remains. 
\end{abstract}


\date{\today}

\maketitle

%%%%%%%%%%%%%%%%%%%%%%%%%%%%%%%%%%%%%%%%%%%%%%%%%%%%%%%%%%%%%%%%%%%%%%%%%%%%
\section{Introduction}

The Kondo problem \cite{Kondo64,HewsonBook}, namely the physics of a
magnetic impurity weakly coupled to a sea of otherwise non-interacting
electrons, is one of the most thoroughly studied questions of
many-body solid state physics. One reason for this ongoing interest is
that the Kondo problem is a deceptively simple model system which
nevertheless displays very non-trivial behavior and so requires the
use of a large variety of theoretical tools to be thoroughly
understood, including exact approaches (the numerical renormalization
group \cite{WilsonRMP75,NRGreviewRMP08}, Bethe ansatz techniques
\cite{Andrei80,Wiegmann80}, and bosonization
\cite{BosonizationForKondoReview,Schotte69,Schotte70,Blume70}) as well
as various approximation schemes (perturbative renormalization
\cite{Anderson70,Fowler71} and mean field theories
\cite{MeanFieldForSpinReview,MeanFieldForSpinearlypaper1,Coleman83,Read84}).

In its original form, the Kondo problem refers to a dilute set of real
magnetic impurities (e.g.\ \textit{Fe}) in some macroscopic metallic
host (say \textit{Au}). In such circumstances, the density of states of
the metallic host can be considered as flat and featureless within the
energy scale at which the Kondo physics takes place. Modeling that
case with a simple impurity model such as either the $s$-$d$ model or the Anderson impurity model \cite{HewsonBook}, one finds that a
single energy scale, the Kondo temperature $\Tk$, emerges and
distinguishes two rather different temperature regimes. For
temperatures $T$ much larger than $\Tk$, the magnetic impurity behaves
as a free moment with an effective coupling which, although
renormalized to a larger value than the (bare) microscopic one,
remains small. For $T \ll \Tk$ on the other hand, the magnetic
impurity is screened by the electron gas and the system behaves as a
Fermi liquid \cite{Nozieres74} characterized by a phase shift and a 
residual interaction associated with virtual breaking of the Kondo
singlet. 

That the Kondo effect is in some circumstances relevant to the physics
of quantum dots was first theoretically predicted
\cite{Ng88,Glazman88} and then considerably later confirmed
experimentally \cite{Goldhaber98,Cronenwett98,VanDerWiel00}. Indeed,
for temperatures much lower than both the mean level spacing and the
charging energy, a small quantum dot in the Coulomb blockade regime
can be described by the Anderson impurity model, with the dots playing
the role of the magnetic impurity and the leads the role of the
electron sea. Quantum dots, however, bring the possibility of two
novel twists to the traditional Kondo problem. The first follows from
the unprecedented control over the shape, parameters, and spatial
organization of quantum dots: such control makes it possible to design
and study more complex ``quantum impurities'' such as the two-channel, 
two impurity, or SU(4) Kondo problems \cite{GoldhaberGRev07,ChangRev09}.
The second twist, which shall be our main concern here, is that the
density of states in the electron sea may have low energy structure
and features, in contrast to the flat band typical of the original
Kondo effect in metals.

Indeed, the small dot playing the role of the quantum impurity need
not be connected to macroscopic leads, but rather may interact instead
with a larger dot. The larger dot may itself be large enough to be
modeled by a sea of non-interacting electrons (perhaps with a constant
charging energy term) but, on the other hand, be small enough to be
fully coherent and display finite size effects
\cite{Kouwenhoven97}. These finite size effects introduce two
additional energy scales into the Kondo problem. The first is simply
the existence of a finite mean level spacing, leading to what has been
called the ``Kondo box'' problem by Thimm and coworkers
\cite{Thimm99}.  The other energy scale introduced by the finite
electron sea is the Thouless energy $E_{\rm Th} = \hbar/\tau_{c}$
where $\tau_c$ is the typical time to travel across the
``electron-reservoir'' dot. When probed with an energy resolution
smaller than $E_{\rm Th}$, both the spectrum and the wave-functions of
the electron sea display mesoscopic fluctuations \cite{Kouwenhoven97}, 
which will affect
the Kondo physics and hence lead to what has been called 
the ``mesoscopic Kondo problem" \cite{KaulEPL05}. 
Similar studies were also conducted in the context of disordered systems
\cite{Kettemann04,Kettemann06}.

Both the Kondo box problem and the high temperature regime of the
mesoscopic Kondo problem are by now reasonably well understood. For a
finite but constant level spacing in the large dot, various
theoretical approaches ranging from the non-crossing approximation
\cite{Thimm99} and slave boson mean field theory \cite{Simon03} to
exact quantum Monte Carlo \cite{Yoo05,KaulPRL06,Kaul09} and numerical
renormalization group methods
\cite{Cornaglia02a,Cornaglia02b,Cornaglia03} have been used to map out
the effect on the spectral function \cite{Cornaglia02b}, persistent
current \cite{Kang00,Affleck01}, conductance
\cite{Simon02,Simon06,Pereira08}, and magnetization
\cite{Cornaglia02a,KaulPRL06,Kaul09}. In the same way, a mix of
perturbative renormalization group analysis \cite{Zarand96,KaulEPL05,Bedrich10}
and quantum Monte Carlo \cite{KaulEPL05} have made it possible to
understand the high temperature regime of the mesoscopic problem (see
also Refs.\,\onlinecite{Kettemann04,Kettemann06,Kettemann07,Zhuravlev08}
for treatment of disordered systems). The picture that emerges is that
mesoscopic fluctuations of the density of states translate into
mesoscopic fluctuations of the Kondo temperature, but that once this
translation has been properly taken into account, the high-temperature
physics remains essentially the same as in the flat band case. In
particular, physical properties can be written as the same universal
function of the ratio $T/\Tk$ as in the bulk flat-band case, as long
as $\Tk$ is understood as a realization dependent parameter
\cite{KaulEPL05}. In this sense, the Kondo temperature remains a
perfectly well defined concept (and quantity) in the mesoscopic
regime, as long as it is defined from the high-temperature behavior.

In contrast, the consequences of mesoscopic fluctuations on Kondo
physics in the low temperature regime, $T \!\ll\! \Tk$, remain largely
unexplored. A few things are nevertheless known: for instance, using
the example of the local susceptibility, exact Monte Carlo
calculations have confirmed that below $\Tk$ physical quantities do
not have the universal character typical of the traditional (flat
band) Kondo problem \cite{KaulEPL05}. This result is not surprising
since the mesoscopic fluctuations existing at all scales between the
mean level spacing $\Delta$ and $E_{\rm Th}$ introduce in some sense a
much larger set of parameters in the definition of the problem,
leaving no particular reason why all physical quantities should be
expressed in terms of $T/\Tk$. Thus, the low temperature regime of the
mesoscopic Kondo problem should display non-trivial but interesting
features. On the other hand, it seems reasonably clear that the very
low temperature regime should be described by a Nozi\`eres-Landau
Fermi liquid, as in the original Kondo problem. Indeed, the physical
reasoning behind the emergence of Fermi liquid behavior at low
temperatures, namely that for energies much lower than $\Tk$ the
impurity spin has to be completely screened, applies as well in the
mesoscopic case as long as $T,\Delta \ll \Tk$.

As a consequence, the mesoscopic Kondo problem provides an interesting
example of a system which, as the temperature is lowered, starts as a
(nearly) non-interacting electron gas with some mesoscopic
fluctuations when $T \gg \Tk$, goes through an intrinsically
correlated regime for $T \simeq \Tk$, and then becomes again a
non-interacting electron gas (essentially) with \emph{a priori}
different mesoscopic fluctuations as $T$ becomes much smaller than
$\Tk$. \emph{A natural question, then, is to characterize the 
correlation between the statistical fluctuations of the electron gas
corresponding to the two limiting regimes.} The goal of this paper
is to address this issue (some preliminary results were reported in Ref.\,\onlinecite{SBMFT_short}). 
As an exact treatment of the low temperature
mesoscopic Kondo problem is not an easy task, we shall tackle
this problem here in a simplified framework, namely the one of slave
boson/fermion mean field theory, within which a complete understanding
can be obtained. We shall furthermore limit our study to the case
where the dynamics in the finite ``electron sea'' reservoir is chaotic, 
and thus the statistical fluctuations of the high temperature Fermi
gas is described by random matrix theory \cite{Bohigas91}.

The structure of this paper is as follows. In Sec.\,\ref{sec:model},
we introduce more formally the mesoscopic Kondo model under study and
describe the mean field approach on which the analysis is
based. Sec.\,\ref{sec:MF} is devoted to the fluctuations of the mean
field parameters. Fluctuations of physical static quantities are
analyzed in Sec.\,\ref{sec:globalphys}.  We then turn in
Sec.\,\ref{sec:spectra} to the study of the spectral fluctuations. For
the resonant level model arising from the mean field treatment, we
give in particular a derivation of the spectral joint distribution
function, as well as a simplified analysis, in the spirit of the
Wigner surmise \cite{Bohigas91}, of some correlation functions
involving the levels of the low and high temperature regimes. Wave
function correlations are then considered in
Sec.\,\ref{sec:WF}. Finally, Sec.\,\ref{sec:conclusion} contains some
discussion and conclusions.

%%%%%%%%%%%%%%%%%%%%%%%%%%%%%%%%%%%%%%%%%%%%%%%%%%%%%%%%%%%%%%%%%%%%%%%%%%%%
\section{Model} 
\label{sec:model}

\subsection{Mesoscopic bath + Anderson impurity}
\label{sec:mesoAnderson}

We investigate the low temperature properties of a mesoscopic
bath of electrons (e.g., a big quantum dot), coupled to a magnetic
impurity (e.g.\ a small quantum dot or a magnetic ion). The
Hamiltonian of the system is
\begin{eqnarray}
\label{GeneralHamiltonian}
H=H_{\rm bath}+H_{\rm imp} \; , 
\end{eqnarray}
where $H_{\rm bath}$ describes the mesoscopic electronic bath and $H_{\rm imp}$ 
describes the interaction between the bath and the local magnetic impurity. 
Here, in a particular realization of this general model, the mesoscopic bath is described by the non-interacting 
(i.e.\ quadratic) Hamiltonian  
\begin{eqnarray}
\label{BathHamiltonian}
H_{\rm bath}\equiv 
\sum_{i, \sigma}(\epsilon_{i}-\mu)c_{i\sigma}^{\dagger}c_{i\sigma}~, 
\end{eqnarray}
where $i=1,\cdots ,N$ indexes the level, $\sigma=\uparrow,\downarrow$
is the spin component, and $\mu$ is the chemical potential.  We assume
that, in $H_{\rm imp}$, the local Coulomb interaction $U n_{d\uparrow}
n_{d\downarrow}$ between $d$-electrons is such that $U= \infty$, so
states with two $d$-electrons on the impurity must be projected
out.  With this constraint implemented, the local impurity term is
taken as
\begin{eqnarray}
\label{HimpAndersonboxU1}
H_{\rm imp}=
V_0\sum_{\sigma}[c_{0\sigma}^{\dagger}d_{\sigma}+
d_{\sigma}^{\dagger}c_{0\sigma}]
+E_{d}\sum_{\sigma}d_{\sigma}^{\dagger}d_{\sigma} \; , 
\end{eqnarray}
where the annihilation and creation operators $d_{\sigma}$ and
$d_{\sigma}^{\dagger}$ act on the states of the impurity (small dot).
The state in the reservoir to which the $d$-electrons couple is
labeled ${\bf r}=0$ with the corresponding operator $c_{0\sigma}$
related to the bath eigenstate operators $c_{i\sigma}$ through
\begin{eqnarray}
\label{Defczero}
c_{0\sigma}&=&\sum_{i=1}^{N}\phi_{i}^{\star}(0)c_{i\sigma}~, 
\end{eqnarray}
where $\phi_i({\bf r}) = \langle{\bf r}| i \rangle $ denotes the
one-body wave functions of the $H_{\rm bath}$. The local normalization
relation $\sum_{i}\vert \phi_{i}(0)\vert^{2}=1$ implies that the
average intensity is $\overline{ |\phi_{i}(0)|^2 } = 1/N$, where
$\overline{(\cdot )}$ denotes the configuration average.  Finally, the
width of the $d$-state, $\Gamma_0$, because of coupling to the
reservoir is given in terms of the mean density of states, $\rho_0$,
by
\begin{equation}
\Gamma_0 \equiv \pi \rho_0 V_0^2 \;, \quad 
\rho_0 \equiv \frac{\overline{ |\phi_{i}(0)|^2 } }{ \Delta} =
\frac{1}{D}
\end{equation}
where $D = N\Delta$ is the bandwidth of the electron bath.

To be in the ``Kondo regime'', some assumptions are  made
about the parameters of the Hamiltonians Eqs.~(\ref{BathHamiltonian}) and
(\ref{HimpAndersonboxU1}).  To start, the dimensionless parameter
obtained as the product of  the Kondo coupling, 
\begin{equation}
\label{eq:JK}
J_K \equiv \frac{2 V_0^2}{|E_d|} \; , 
\end{equation}
and the local density of states, $\rho_0$, should be assumed small:
$\rho_0 J_K \ll 1$, or equivalently $\Gamma_0/E_d \ll 1$.  Indeed,
this condition implies that the strength $V_0^2/NE_d$ of the second
order processes involving an empty-impurity virtual-state is much
smaller than the mean level spacing $\Delta$.  Furthermore, as we
discuss in more detail in Section \ref{sec:MF}, the Kondo regime is
characterized by $\Tk \ll \Gamma_0$, for which the fluctuations of the
number of particles on the impurity is weak.  If $\Tk$ increases to
the point that $\Tk \alt \Gamma_0$, one enters the mixed valence
regime where these fluctuations become important.

\subsection{Random matrix model}
\label{sec:RMT}

To study the mesoscopic fluctuations of our impurity model, we assume 
chaotic motion in the reservoir in the classical limit. Random matrix theory (RMT)
provides a good model of the quantum energy levels and wave functions in this
situation \cite{Bohigas91,Kouwenhoven97}: we use the Gaussian orthogonal
ensemble (GOE, $\beta=1$) for time reversal symmetric systems and the
Gaussian unitary ensemble (GUE, $\beta=2$) for non-symmetric systems
\cite{MehtaBook,Bohigas91}.  The joint distribution function of the
unperturbed reservoir-dot energy levels is therefore given by
\cite{MehtaBook}
\begin{equation} \label{eq:WD_dist}
P_\beta(\epsilon_1,\epsilon_2,\cdots,\epsilon_N) 
\propto
\prod_{i>j} \left|\epsilon_i - \epsilon_j\right|^\beta
\exp\left(-\frac{1}{4\alpha^2} \sum_i \epsilon_i^2 \right)~, 
\end{equation}
(with $\alpha = \sqrt{N}\Delta/\pi$ where $\Delta$ is the mean
level spacing in the center of the semicircle).  The corresponding
distribution of values of the wave function at ${\bf r}=0$, the site
in the reservoir to which the impurity is connected, is the
Porter-Thomas distribution,
\begin{equation} \label{eq:PT}
p_\beta(x_i=N|\phi_i(0)|^2) = \frac{1}{(2\pi x_i)^{1-\beta/2}}
  \exp \left( -\frac{\beta}{2} x_i \right) \; .
\end{equation}
Furthermore, in the GOE and GUE, the eigenvalues and eigenvectors are
uncorrelated. 

For the GOE and GUE, the mean density of
states follows a semicircular law---a result that is rather unphysical.  Except
when explicitly specified, we assume either that  we consider only
the center of that the semicircle or some rectification procedure has
been applied, so  we effectively work with a flat mean density of
states.

\subsection{Slave boson mean-field approximation}
\label{sec:SBMFT}

Following the standard procedure
\cite{Coleman83,Read84,MeanFieldForSpinReview,MeanFieldForSpinReview2,MeanFieldForSpinReview3,BurdinRev2009},
we introduce auxiliary boson $b^{(\dagger )}$ and fermion
$f_{\sigma}^{(\dagger )}$ annihilation (creation) operators, such that
$d_{\sigma} = b^{\dagger}f_{\sigma}$, with the constraint
\begin{eqnarray}
\label{ConstraintAndersonU1}
b^\dagger b+\sum_{\sigma}f^{\dagger}_{\sigma}f_{\sigma}=1~. 
\end{eqnarray}
The impurity interaction~(\ref{HimpAndersonboxU1}) is  rewritten as 
\begin{eqnarray}
\label{HimpAndersonboxU2}
H_{\rm imp}=
V_0\sum_{\sigma}[b^{\dagger}c_{0\sigma}^{\dagger}f_{\sigma}+
bf_{\sigma}^{\dagger}c_{0\sigma}]
+E_{d}\sum_{\sigma}f_{\sigma}^{\dagger}f_{\sigma}~. 
\end{eqnarray}
The mapping between physical states and auxiliary states of the impurity 
is 
\begin{eqnarray}
{\rm Physical~state} &\rightarrow& {\rm Auxiliary~state}\nonumber\\
\vert  \emptyset \rangle~~~~~ 
&\rightarrow& ~~~~~b^{\dagger}\vert \emptyset \rangle \nonumber  \\ 
\vert \sigma\rangle~~~~~
&\rightarrow& ~~~~~f_{\sigma}^{\dagger}\vert \emptyset \rangle 
\qquad \qquad . \nonumber\\
\vert \uparrow\downarrow\rangle~~~~~
&\rightarrow& {\rm projected~out}\nonumber \;.
\end{eqnarray}
This auxiliary operator representation is exact in the limit $U=\infty$ 
as long as the constraint~(\ref{ConstraintAndersonU1}) is satisfied 
and the bosonic term in $H_{\rm imp}$ is treated exactly. 

Note that we use here a slave boson formalism with U(1) gauge
symmetry.  Generalized slave boson fields have been introduced in
order to preserve the SU(2) symmetry of the model, as discussed in
Refs.\,\onlinecite{MeanFieldForSpinReview} and \onlinecite{Affleck88}.
Such a generalized SU(2) slave boson approach would not change crucially
the physics of the single impurity mean-field solution, but it may
become relevent for models with more than one impurity.

The mean-field treatment of the Anderson box Hamiltonian invokes two 
complementary approximations: 
{\it (i)}~The bosonic operator $b$ is considered a complex field, with 
an amplitude $\eta$ and a phase $\theta$. Since the Hamiltonian is invariant 
with respect to the $U(1)$ gauge transformation $b\to be^{i\theta}$ and 
$f_{\sigma}\to f_{\sigma}e^{i\theta}$, the phase $\theta$ is not 
a physical observable, and we choose $\theta=0$: 
\begin{eqnarray}
\label{MFAndersonUeta1}
b,~b^{\dagger}&\mapsto& \eta~, 
\end{eqnarray}
where $\eta$ is a positive real number. This approximation 
corresponds to assuming that the bosonic field condenses. 
{\it (ii)}~The constraint~(\ref{ConstraintAndersonU1}) is satisfied on 
average, by introducing a static Lagrange multiplier, $\xi$. 
The Hamiltonian of Eq.~(\ref{GeneralHamiltonian}) 
treated within the slave boson mean-field approximation thus reads  
\begin{widetext}
\begin{eqnarray}
\label{MFHamiltonianAndersonbox1}
H_{\rm MF} =\sum_{\sigma}\left( 
\sum_{i=1}^{N}\left[ 
(\epsilon_{i}-\mu)c_{i\sigma}^{\dagger}c_{i\sigma}
+
\eta V_0 \phi_{i}^{\star}(0)f_{\sigma}^{\dagger}c_{i\sigma}
+
\eta V_0\phi_{i}(0)c_{i\sigma}^{\dagger}f_{\sigma}
\right]
+(E_{d}-\xi)f_{\sigma}^{\dagger}f_{\sigma}\right)
+\xi(1-\eta^{2}) \; . 
\end{eqnarray}
\end{widetext}
The mean-field parameters $\eta$ and $\xi$ must be chosen to minimize
the free energy of the system, ${\cal F}=- T \ln({{\rm Tr} [e^{-H_{\rm
      MF}/T}])}$, yielding the saddle point relations
\begin{eqnarray}
\label{Consistenteta1}
2\eta\xi
&=&
V_0\sum_{\sigma}\left[ 
\langle f_{\sigma}^{\dagger}c_{0\sigma}\rangle
+ \langle c_{0\sigma}^{\dagger}f_{\sigma}\rangle 
\right] \quad~, \\
\label{Consistentxi1}
1-\eta^{2}
&=&
\sum_{\sigma}\langle f_{\sigma}^{\dagger}f_{\sigma}\rangle ~, 
\end{eqnarray}
where the thermal averages $\langle\cdots\rangle$ have to be computed 
self-consistently from the mean-field Hamiltonian.
\footnote{The chemical potential $\mu$ can be considered as an external tunable 
parameter, or determined self-consistently if one considers a given electronic 
occupancy $N_{c}$. In this latter case, 
Eqs.~(\ref{Consistenteta1})-(\ref{Consistentxi1}) have to be completed by a 
third relation: 
\protect{$N_{c} = \sum_{l\sigma} \langle c_{l\sigma}^{\dagger}
c_{l\sigma}\rangle$}.
}

\subsection{Method for solving the mean-field equations} 

In this section, we explain how to solve the self-consistent equations
for the effective parameters, $\eta$ and $\xi$. We start by
introducing the imaginary-time equilibrium Green functions
\begin{eqnarray}
G_{ff}(\tau-\tau')&\equiv 
\langle\langle f_\sigma(\tau);f_\sigma^{\dagger}(\tau')\rangle\rangle~, \\
G_{fi}(\tau-\tau')&\equiv 
\langle\langle f_\sigma(\tau);c_{i\sigma}^{\dagger}(\tau')\rangle\rangle~, \\
G_{if}(\tau-\tau')&\equiv 
\langle\langle c_{i\sigma}(\tau);f_\sigma^{\dagger}(\tau')\rangle\rangle~, \\
G_{ij}(\tau-\tau')&\equiv 
\langle\langle c_{i\sigma}(\tau);c_{j\sigma}^{\dagger}(\tau')\rangle\rangle~. 
\end{eqnarray}
Using the equations of motion from the mean-field
Hamiltonian Eq.\,(\ref{MFHamiltonianAndersonbox1}) and after
straightforward algebra, we find
\begin{eqnarray}
\label{AndersonboxGff1}
G_{ff}(i\omega_{n})
&=&\!
\left[ i\omega_{n}+\xi -E_{d}-\eta^{2}V_0^{2}
\sum_{i=1}^{N}\frac{\vert \phi_i(0)\vert^{2}}{i\omega_{n}+\mu-\epsilon_{i}}
\right]^{-1}  \nonumber\\
&&\\
\label{AndersonboxGlf1}
G_{if}(i\omega_{n})
&=&
\frac{\eta V_0\phi_{i}(0)}{i\omega_{n}+\mu-\epsilon_{i}}
G_{ff}(i\omega_{n}) \; , \\
&&\nonumber\\
\label{AndersonboxGfl1}
G_{fi}(i\omega_{n})
&=&
\frac{\eta V_0\phi_{i}^{\star}(0)}{i\omega_{n}+\mu-\epsilon_{i}}
G_{ff}(i\omega_{n}) \; , \\
&&\nonumber\\
\label{AndersonboxGll1}
G_{ij}(i\omega_{n})
&=& 
\frac{\delta_{ij}}{i\omega_{n}+\mu-\epsilon_{i}} \; + \\
&& 
\frac{\eta V_0 \phi_{j}^{\star}(0)}{i\omega_{n}+\mu-\epsilon_{j}}
G_{ff}(i\omega_{n})
\frac{\eta V_0 \phi_{i}(0)}{i\omega_{n}+\mu-\epsilon_{i}} 
\; ,  \nonumber
\end{eqnarray}
where 
$\omega_{n}\equiv (2n+1)\pi T$ are the fermionic Matsubara frequencies.
Finally, the mean-field equations~(\ref{Consistenteta1})-(\ref{Consistentxi1}) 
for $\eta$ and $\xi$ can be rewritten as 
\begin{eqnarray}
\label{Consistenteta3}
\eta\xi&=& V_0T\sum_{i=1}^{N}
\sum_{n=-\infty}^{+\infty}
[\phi_{i}^{\star}(0)G_{if}(i\omega_{n})
+\phi_{i}(0)G_{fi}(i\omega_{n})]~, \nonumber\\
&&\\
\label{Consistentxi3}
1&=&
\eta^{2}+
2T
\sum_{n=-\infty}^{+\infty}
G_{ff}(i\omega_{n}) \; .
\end{eqnarray}
Self-consistency therefore can be achieved by iterating successively 
Eqs.~(\ref{AndersonboxGff1})-(\ref{AndersonboxGfl1}), which define the
Green functions in terms of the parameters $\xi$ and $\eta$, and
Eqs.~(\ref{Consistenteta3})-(\ref{Consistentxi3}), which fix $\xi$ and
$\eta$ from the Green functions.

\begin{figure}[t]
\centering
\includegraphics[width=3.3in,clip]{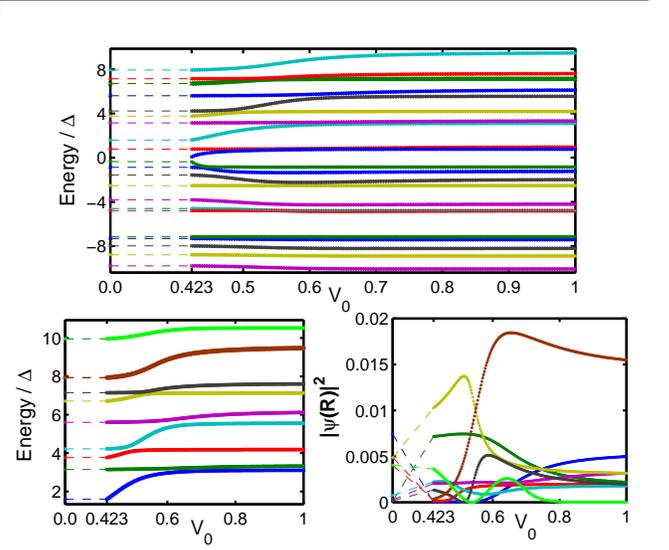}
\vspace*{-0.1in}
\caption{(Color online) The evolution of the energy levels and wave
  functions as a function of coupling strength for a realization drawn
  from the GOE (calculated using infinite-$U$ SBMFT). (a)~Energy
  levels near the Fermi energy $\mu=0$ for coupling $V_0$ (from
  0.423 to 1.0). (b)~Zoom of a few levels above the Fermi
  energy. (c)~The wave function amplitudes $|\psi({\bf R})|^2$
  corresponding to the energy levels in (b) for an arbitrary position
  ${\bf R} \neq 0$. Parameters: band width $D=3.$, $E_d=-0.7$,
  $\Delta=0.0075$, and  $T=0.005$. 
}
\label{fig:EW_GOE}
\end{figure}

\begin{figure}[t]
\centering
\includegraphics[width=3.45in,clip]{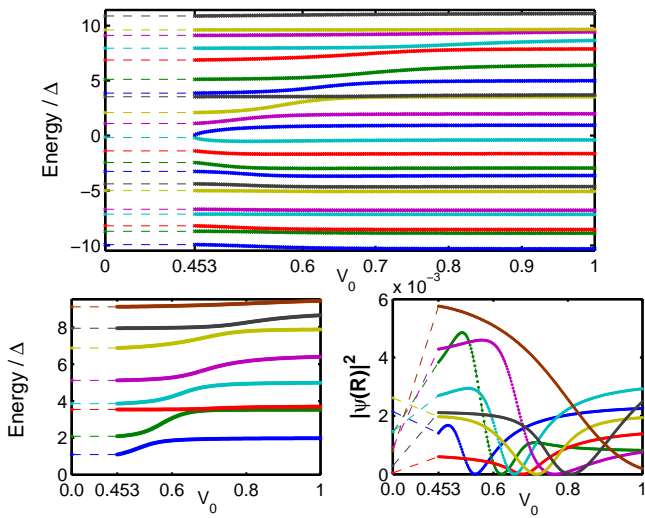}
\vspace*{-0.2in}
\caption{(Color online) The analog of Fig.~\ref{fig:EW_GOE} for the GUE; parameters are the same. Note that the variation is smoother for these GUE results than for the GOE in Fig.~\ref{fig:EW_GOE}.
}
\label{fig:EW_GUE}
\end{figure}

As an example of the output from this procedure, we show in
Figs.~\ref{fig:EW_GOE} and \ref{fig:EW_GUE}, as a function of the
strength of the coupling $V_0$, the one-body energy levels that result
from a slave-boson mean field theory (SBMFT) treatment of the Anderson
box for a particular realization of the box. As we discuss in more
detail below (see section \ref{sec:JKc}), a non-trivial solution of
the SBMFT equations exists only for $J_K$ above some critical value
$J^c_K$, or equivalently [see Eq.~(\ref{eq:JK})] for $V_0$ larger than
a threshold $V^c_0$.  We thus show the non-interacting levels below
that value and break the axis at that point [$V^c_0 \simeq 0.423$
(GOE) and $V^c_0 \simeq 0.453$ (GUE) for the realizations chosen].
Clearly, the levels do indeed shift substantially as a function of
coupling strength; notice as well the additional level injected near
the Fermi energy. The change in the levels occurs more sharply and for
slightly smaller values of $V_0$ in the GOE case than for the
GUE. Finally, we observe that, as one follows a level as a function of
$V_0$, little change occurs after some point.  The coupling strength
$V_0$ at which levels reach their limiting value depends on the
distance to the Fermi energy; it corresponds to the point where the
Abrikosov-Suhl resonance becomes large enough to include the
considered level. These limiting values of the energies are the SBMFT
approximation to the single quasi-particle levels of the Nozi\`eres
Fermi liquid theory.

\subsection{Qualitative behavior}
\label{sec:GenFeat}

Before entering into the detailed quantitative analysis, we describe
here some simple general properties of the mesoscopic Kondo problem
within the SBMFT perspective.

We note first that the mean-field
equations~(\ref{Consistenteta3})-(\ref{Consistentxi3}), have a trivial
solution $\eta=0$ and $\xi=E_{d}$.  This solution is actually the
only one in the high temperature regime: the
mesoscopic bath is effectively decoupled from the local magnetic
impurity which can be considered a free spin-$1/2$.  The onset
of a solution $\eta\neq 0$ defines, in the mean field approach, the
Kondo temperature $\Tk$.

Below $\Tk$, the self-consistent mean-field approach results in an
effective one-particle problem, specifically a resonant level model
with resonant energy $E_d\!+\!\mu\! -\! \xi$ and effective coupling
$\eta V_0$.  This resonance is interpreted as the Abrikosov-Suhl
resonance characterizing the one-particle local energy spectrum of the
Kondo problem below $\Tk$. The width of this resonance, $\Gamma(\eta)
= \eta^2 \Gamma_0$, vanishes for $T=\Tk$, and quickly reaches a value
of order $\Tk$ when $T \ll \Tk$ (more detailed analysis is in
Sec.\,\ref{sec:qualitativdescription}).  Note that the mesoscopic
Kondo problem differs from the bulk case: mesoscopic fluctuations may
affect the large but finite number of energy levels that lie within
the resonance.

The Anderson box is, however, a many-body problem.  Its ground state
cannot be described too naively in terms of one-body electronic wave
functions, and more generally one should question the validity of the
one particle description for each physical quantity under
investigation.  In this respect however, the configuration we
consider, namely the low temperature regime of the Kondo box problem,
is particularly favorable.  Indeed, the line of argument developed by
Nozi\`eres \cite{Nozieres74} to show that the low temperature regime
of the Kondo problem is a Fermi-liquid applies equally well in the
mesoscopic case as in the bulk one for which it was originally
devised.  Therefore, as long as both the temperature $T$ and the mean
level spacing $\Delta$ are much smaller than the Kondo temperature, we
\emph{a priori} expect the physics of the Kondo box to be described in
terms of fermionic quasi-particles.  The notions of one particle
energies and wavefunction fluctuations in the strong interaction
regime, which will be our main concern below, are therefore
relevant.  We take the point of view that, as in the bulk case
\cite{MeanFieldForSpinReview,BurdinRev2009,Coleman83,Read84}, the
mean field approach provides a good approximation for these
quasi-particles in this low temperature regime, and therefore for the
physical quantities derived from them \cite{Ullmo2011}.  As we shall
see furthermore, most of the fluctuation properties we shall
investigate have universal features that makes them largely
independent from possible corrections to this approximation (such as,
for instance, corrections on the Kondo temperature), making the 
approach we are following particularly robust.

%%%%%%%%%%%%%%%%%%%%%%%%%%%%%%%%%%%%%%%%%%%%%%%%%%%%%%%%%%%%%%%%%%%%%%%%%%%%

\section{Fluctuations of the mean field parameters}
\label{sec:MF}

To begin our investigations of the low temperature properties of the
mesoscopic Kondo problem within SBMFT, we
consider the fluctuations of the mean field parameters $\eta$ and
$\xi$ appearing in Eq.~(\ref{MFHamiltonianAndersonbox1}).  We shall
comment also on the degree to which these fluctuations are connected with
those of the Kondo temperature
$\Tk$ \cite{Kettemann04,KaulEPL05,Bedrich10}.

\subsection{Preliminary analysis}
\label{sec:preliminary}

We start with a few basic comments about the
eigenvalues $\{\lambda_\kappa-\mu\}$ and eigenstates $|\psi_\kappa\rangle$
($\kappa = 0,1,\cdots,N$) of the mean field Hamiltonian
Eq.~(\ref{MFHamiltonianAndersonbox1}).  Concerning
the latter, we shall be interested in the two quantities, 
\begin{eqnarray}
u_\kappa  & \equiv & |\langle f| \psi_\kappa\rangle |^2~,  \label{eq:uk_def}\\
\theta_\kappa & \equiv &  \langle 0 | \psi_\kappa\rangle \langle 
\psi_\kappa  | f\rangle \label{eq:thetak_def} \; .
\end{eqnarray}
$u_\kappa$ measures the overlap probability between the eigenstate
$\kappa$ and the impurity state $| f\rangle$, and $\theta_\kappa$ the
admixture of this eigensate with $| f\rangle$ and $|0\rangle= \sum_i
\phi_i (0) | i \rangle$, the electron-bath state connected to the
impurity.  Note that $\theta_\kappa$ is a real quantity. In this
section we use $\kappa =0$ to denote the additional resonant
level added to the original system, and so in the limit $V_0 \to 0$,
one has $|\psi_0\rangle \to |f\rangle$ and $\lambda_\kappa \to
  \epsilon_{i=\kappa}$ ($\kappa=1,\ldots,N$).

Expressing the Green function of the mean field Hamiltonian  as
\begin{equation} \label{eq:Green}
\hat G(\lambda-\mu) = [\lambda-\mu-H_{\rm MF}]^{-1}
=
\sum_{\kappa=0}^N \frac{|\psi_\kappa\rangle \langle
  \psi_\kappa|}{\lambda - \lambda_\kappa} \; ,
\end{equation}
we can check that $(\lambda_\kappa-\mu)$  are the poles  of the Green function 
$G_{ff}(z) =\langle f | \hat G(z) |f\rangle$.  From
Eq.~(\ref{AndersonboxGff1}) we have therefore immediately that the
$\lambda_\kappa$ are the solutions of the equations
\begin{equation} \label{eq:lambdas}
\frac{\Delta}{\pi}  \sum_{i=1}^{N}\frac{x_i}{\lambda-\epsilon_{i}} 
=
\frac{\lambda - \mathcal{E}_0(\xi)}{\Gamma(\eta)} 
\; ,
\end{equation}
where we have used the notation
\begin{eqnarray}
\mathcal{E}_0(\xi) & \equiv&  E_d + \mu - \xi \nonumber\\
\Gamma(\eta) & \equiv &  \eta^2 \Gamma_0 = \pi \rho_0 \eta^2 V_0^2
\end{eqnarray}
for the center and the width of the resonance, and $x_i \equiv N \vert
\phi_i(0)\vert^{2}$ for the normalized wavefunction probability at
${\bf r}=0$.  Note first that Eq.~(\ref{eq:lambdas}) implies that
there is one and only one $\lambda_\kappa$ in each interval
$[\epsilon_i,\epsilon_{i+1}]$: the two sets of eigenvalues are
interleaved and so certainly heavily correlated. Furthermore,
$|\langle f |\psi_\kappa\rangle |^2$ are the corresponding residues,
so  again, from Eq.~(\ref{AndersonboxGff1}),
\begin{equation} \label{eq:uk}
u_\kappa = \frac{1}{1 +
\displaystyle
  \frac{\Gamma(\eta)}{\pi} \sum_{i=1}^N \frac{x_i
    \Delta}{(\lambda_\kappa - \epsilon_i)^2}} \; . 
\end{equation}

Eq.~(\ref{eq:lambdas}) is easily solved outside of the resonance,
i.e.\ when $|\lambda -\mathcal{E}_0(\xi)| \gg \Gamma(\eta)$: in
that case one contribution $i(\kappa)$ dominates the sum on the left
hand side. [With our convention where $\kappa=0$ corresponds to
  the extra level added to the original system, we actually just have
  $i(\kappa) = \kappa$.] The solution for the fractional shift in
the level, $\delta_\kappa \equiv (\lambda_\kappa -
\epsilon_{i(\kappa)}) / \Delta$ is then given by
\begin{equation} \label{eq:deltak_out}
  \delta_\kappa \simeq \frac{\Gamma(\eta)}{\pi}
  \frac{x_{i(\kappa)}}{\lambda_\kappa - \mathcal{E}_0} \ll 1 \; . 
\end{equation}
Eq.~(\ref{eq:uk}) and (\ref{eq:deltak_out}) then yield for the wave
function intensity 
\begin{equation} \label{eq:uk_out}
u_\kappa \simeq  \frac{\Gamma(\eta)}{\pi} 
\frac{x_{i(\kappa)} \Delta}{(\lambda_\kappa  - \mathcal{E}_0)^2} 
 \ll \frac{\Delta}{\Gamma(\eta)} \; .  
\end{equation}

If the resonance is small [$\Gamma(\eta) \ll \Delta$], all states are
accounted for in this way, except for $\lambda_0 \simeq \mathcal{E}_0$
which is then such that $u_{\kappa=0} \simeq 1$.

If the resonance is large, $\Gamma(\eta) \gg \Delta$, the states
within the resonance -- those satisfying $|\lambda
-\mathcal{E}_0(\xi)| \ll \Gamma(\eta)$ -- must be treated
differently. Because the left hand side of Eq.\,(\ref{eq:lambdas}) can
be neglected in this regime, these states have only a weak dependence
on $\Gamma(\eta)$.  The typical distance between a $\lambda_\kappa$
and the closest $\epsilon_i$ is then of order $\Delta$, and the
corresponding wave functions participate approximately equally in the
Kondo state,
\begin{equation}\label{eq:uk_in}
u_\kappa \sim \Delta/\Gamma(\eta) \quad \mbox{[inside the resonance]} \; .
\end{equation}

In a similar way, the admixture coefficient, $\theta_\kappa$, is the residue of
$\langle 0 | \hat G(z) | f\rangle = \sum_i \phi^\star (0)
G_{if} (z)$ at the pole $z_\kappa=\lambda_\kappa-\mu$. Applying
Eqs.\,(\ref{AndersonboxGlf1}) and (\ref{eq:lambdas}),
we thus immediately have
\begin{equation} \label{eq:thetak2}
\theta_\kappa =  u_\kappa \cdot \frac{1}{ \eta V_0}
\frac{\Gamma(\eta) \Delta}{\pi} \sum_i
\frac{x_i}{\lambda_\kappa - 
  \epsilon_i}
=  u_\kappa \cdot \frac{\lambda_\kappa - \mathcal{E}_0}{\eta V_0} \; .
\end{equation}
Assuming the resonance is large [$\Gamma(\eta) \gg \Delta$], and
inserting the limiting behaviors of $u_\kappa$ 
Eqs.~(\ref{eq:uk_out})-(\ref{eq:uk_in}), we obtain
\begin{eqnarray}  \label{eq:thetaout}
\theta_\kappa\simeq \theta^{\rm out}_\kappa & = & \frac{x_{i(\kappa)}}{\pi}
\frac{\Delta}{\eta V_0}  \frac{\Gamma(\eta)}{\lambda_k -   \mathcal{E}_0} 
\quad  [|\lambda_\kappa\! -\! \mathcal{E}_0| \gg \Gamma]~,  \\
\label{eq:thetain}
\theta_\kappa & \sim & \frac{\Delta}{\eta V_0}  \frac{\lambda_k -
  \mathcal{E}_0}{\Gamma(\eta)} 
\qquad  [|\lambda_\kappa \!- \!\mathcal{E}_0| \ll \Gamma] \; .
\end{eqnarray}

\subsection{Formation of  the resonance}
\label{sec:qualitativdescription}

Before considering the fluctuations of the mean field parameters
$\eta$ and $\xi$ , let us first discuss the physical mechanisms that
determine their value.  While this discussion is not specific to the
mesoscopic Kondo problem, it is useful to review it briefly before
addressing the mesoscopic aspects.

The self-consistent equations
(\ref{Consistenteta1})-(\ref{Consistentxi1})  or
(\ref{Consistenteta3})-(\ref{Consistentxi3})
can be written as (performing the summation over Matsubara frequencies
in the standard way \cite{Fetter&Walecka} in the latter case),   
\begin{eqnarray}
\label{Consistenteta4}
2 V_0 \sum_{\kappa=0}^{N} (f_\kappa \! - \! \frac{1}{2}) \,
\theta_\kappa  &=& \eta\xi \; ,
\\
\label{Consistentxi4}
n_f = \sum_{\kappa = 0}^N f_\kappa
u_\kappa &=&   \frac{1 - \eta^2}{2} \; ,
\end{eqnarray}
where $f_\kappa = f(\lambda_\kappa \! - \! \mu) =
[1+\exp((\lambda_\kappa \! - \! \mu)/T)]^{-1}$ is the Fermi occupation
number.  One furthermore has the sum rules $\sum_\kappa u_\kappa =
\langle f|f \rangle = 1$ and $\sum_\kappa \theta_\kappa =
\langle 0|f \rangle = 0$ [the latter has been used to generate the
1/2 in (\ref{Consistenteta4})].

As mentioned in Sec.\,\ref{sec:GenFeat}, the trivial solution
of these mean-field
equations ($\eta\!=\!0,\xi\!=\!E_{d}$) is the only one
in the high temperature regime. The Kondo temperature $\Tk$ is
defined, in the mean field approach, as the highest temperature for
which a $\eta\neq 0$ solution occurs. One obtains an equation for 
$\Tk$ by requiring that the non-trivial solution of the mean-field
equations continuously vanishes, $\eta \to 0^{+}$, in which case
$\lambda_{\kappa=0} \to \mathcal{E}_0(\xi)$,
$u_{\kappa=0} \to 1$, and $u_{\kappa \neq 0} \to 0$.
Eq.~(\ref{Consistentxi4}) then reduces to $f(\mathcal{E}_0(\xi) -
\mu) = 1/2$, implying $\mathcal{E}_0(\xi) =
\mu$ and so $\xi = E_d$. Using Eq.~(\ref{eq:thetaout}) to simplify Eq.~(\ref{Consistenteta4}) then gives the mesoscopic version \cite{Bedrich10} of 
the Nagaoka-Suhl equation
\cite{Nagaoka65,Suhl65}
\begin{equation}
\label{eq:TKAndersonbox2}
\frac{E_{d}}{V_0^{2}}
= \sum_{i=1}^{N}
\frac{\vert \phi_{i}(0)\vert^{2}}{\epsilon_{i}-\mu}
\tanh{[(\epsilon_{i}-\mu)/2\Tk]} \; . 
\end{equation}
The same equation for $\Tk$ was obtained from a one-loop
perturbative renormalization group treatment \cite{Zarand96,KaulEPL05}.

In the bulk limit ($N\to\infty$ and no fluctuations) and for $\mu$ in
the middle of the band, this gives  $\Tk^{\rm
  bulk} = a_K (D/2) e^{-1/J_K\rho_0}$ for the Kondo temperature, with $a_K \simeq
1.13\cdot\cdot$ as shown in Appendix~\ref{app:bulk}.  Unless explicitly
specified, we will always assume this quantity is large compared to the
mean level spacing. In this case, the fluctuations of the
Kondo temperature for chaotic dynamics described by the random
matrix model in Sec.\,\ref{sec:RMT} has been analyzed in
Refs.\,\onlinecite{KaulEPL05}-\onlinecite{Kettemann04} and more
recently using SBMFT in
Ref.\,\onlinecite{Bedrich10}.  The main result is that $\delta \Tk$, 
the fluctuation of the Kondo
temperature around the bulk Kondo temperature, scales as
\begin{equation} \label{eq:Tk_fluct}
 \overline{(\delta \Tk)^2 } \sim \Tk^{\rm bulk}\Delta \; .
\end{equation}

Now consider what happens as $T$ decreases further
below $\Tk$.  Dividing Eq.~(\ref{Consistenteta4}) by $\eta V_0$, we
can write it as 
\begin{equation}
\label{Consistenteta5}
\frac{\xi}{V_0^{2}}
= \sum_{\kappa=0}^{N}
 r_\kappa \frac{\vert \phi_{i(\kappa)}(0)\vert^{2}}{\lambda_\kappa - \mathcal{E}_0}
\tanh{[(\lambda_\kappa-\mu)/2T]} \;,
\end{equation}
where $r_\kappa = {\theta_\kappa}/{\theta^{\rm out}_\kappa}$ is one
outside the resonance and scales as $(\lambda_\kappa -
\mathcal{E}_0)^2/\Gamma(\eta)^2$ within the resonance [see
Eqs.~(\ref{eq:thetaout})-(\ref{eq:thetain})]. 
Eq.~(\ref{Consistenteta5}) has a structure very similar to 
the equation for $\Tk$, 
Eq.~(\ref{eq:TKAndersonbox2}).  Indeed, $\xi$ might not be strictly
equal to $E_d$ [and thus $\mathcal{E}_0(\xi)$ might differ slightly
from $\mu$] but its scale will remain the same.  
Then outside the resonance, $\lambda_\kappa \simeq \epsilon_{i(\kappa)}$ 
and $r_\kappa \simeq 1$.  The main difference in the expression 
for $\xi$ is that the logarithmic divergence associated
with the summation of $1/(\lambda_\kappa - \mathcal{E}_0)$ is cutoff
not only by the temperature factor $\tanh{[(\lambda_\kappa-\mu)/2T]}$
at the scale $T$, but also by the ratio $r_\kappa$ at the scale
$\Gamma(\eta)$.  As $T$ becomes significantly smaller than $\Tk$, the
temperature cutoff becomes inoperative. This implies in particular that
$\Gamma(\eta)$ 
will rather quickly switch from 0 to its zero temperature limit when
$T$ goes below $\Tk$.  We shall in the following not consider the
temperature dependence of $\Gamma(\eta)$ but rather focus on its low
temperature limit.  

We see, then, that both $\Tk$ and $\Gamma(\eta)$ represent physically
the scale at which the logarithmic divergence of $\sum_i
|\phi_i(0)|^2/(\epsilon_i - \mu)$ should be cut to keep this sum equal
to $E_d/V_0^2$.  Thus, as long as we are only interested in energy
scales, we can write that for $T \ll \Tk$,
\begin{equation} \label{eq:GammmaScale}
  \Gamma(\eta) \sim \Tk \; .
\end{equation}
The energy dependence of the cutoff $r_\kappa$ within the resonance,
however, differs slightly from that of
$\tanh{[(\lambda_\kappa-\mu)/2\Tk]}$ below $\Tk$. As an
exponentiation is involved, the prefactors of $\Gamma(\eta)$ and $\Tk$ 
somewhat differ; a discussion of the ratio $\Gamma(\eta)/\Tk$ for
the bulk case is given in Appendix~\ref{app:bulk}.

At low temperature, $\eta$ is fixed in such a way that
$\Gamma(\eta)$ is of the scale of the Kondo temperature.  The
condition Eq.~(\ref{Consistentxi4}) then fixes $\xi$, which governs the
center of the resonance $\mathcal{E}_0(\xi)$ so  a
proportion $(1-\eta^2)/2$ of the resonance is below the Fermi energy
$\mu$.  In the  Kondo regime when $n_f \simeq 1/2$,
$\mathcal{E}_0(\xi)$ will therefore remain near $\mu$. In the mixed
valence regime 
$\mathcal{E}_0(\xi)$ will float a bit above $\mu$ for a distance
$- \delta \xi = E_d - \xi$ which scales as 
$\delta \xi \sim \eta^2 \Gamma_0 \alt E_d$.
The order of magnitude of $\xi$ remains thus $E_d$
[as we have assumed above when discussing Eq.~(\ref{Consistenteta5})].

\subsection{Fluctuations scale of the mean field parameters}
\label{sec:FluctScale}

With this physical picture of how the mean field parameters $\eta$ and
$\xi$ are fixed, it is now relatively straightforward to evaluate the
scale of their fluctuations. For simplicity, we assume $T=0$ so 
the mean-field equations become
\begin{eqnarray}
\label{Consistenteta6}
I(\eta,\xi) & \equiv & \pi \sum_{\kappa=0}^{N} 
{\rm sgn}(\lambda_\kappa \!-\!\mu) \frac{(\lambda_\kappa \!
  - \! \mathcal{E}_0)}{\Gamma (\eta)}  \, u_\kappa
  =  \frac{\pi\xi}{\Gamma_0} \quad~, 
\\
\label{Consistentxi6}
 J(\eta,\xi) &\equiv&    \sum_{\lambda_\kappa < \mu} 
u_\kappa =   \frac{1 - \eta^2}{2} \; .
\end{eqnarray}
The discussion below generalizes easily
to finite $T$ as long as it is much smaller than $\Tk$.

The average values of $I(\eta,\xi)$ and $J(\eta,\xi)$ are well
approximated by their ``bulk-value'' analogues $I^{\rm bulk}(\eta,\xi)$
and $J^{\rm bulk}(\eta,\xi)$, obtained with the same global parameters
but with the fluctuating wave-function probabilities $x_i$ replaced by $1$
and the spacing between successive levels taken constant,
$\epsilon_{i+1} \!- \epsilon_{i} \equiv \Delta$. 
We furthermore denote by $(\bar \eta,\bar \xi)$ the solution of
Eqs.~(\ref{Consistenteta6})-(\ref{Consistentxi6}) with $I(\eta,\xi)$
and $J(\eta,\xi)$ replaced by their bulk approximation,  by $\delta
\eta \equiv \eta -  \bar \eta$ and $\delta \xi \equiv \xi \!- \! \bar \xi$ the
fluctuating part of the mean field parameters, and by $\delta I(\eta,\xi)
\equiv I(\eta,\xi) \!- \!I(\eta,\xi)^{\rm bulk}$
and $\delta J(\eta,\xi) \equiv J(\eta,\xi) \!- \! J(\eta,\xi)^{\rm
  bulk}$ the fluctuating parts of the sums appearing in
Eqs.~(\ref{Consistenteta6})-(\ref{Consistentxi6}). 

We start by discussing the  Kondo limit
$\Tk\ll \Gamma_0$, in which case $\bar \eta \ll 1$, 
$\bar \xi \!-\!  E_d \ll \Gamma$, and 
$\bar \Gamma \equiv \Gamma(\bar \eta) = (D/2) \exp(-1/J_K \rho_0)$. 
A calculation in Appendix~\ref{app:bulk} shows
\begin{eqnarray}
I^{\rm bulk}(\eta,\xi) 
 & = & 2 \ln \left(  \frac{D}{2  \Gamma(\eta)} \right) + O(\eta^4)~, 
\label{eq:Ibulk} \\
J^{\rm bulk}(\eta,\xi) 
 & = & \frac{1}{2} + \frac{1}{\pi}
\frac{ \xi \!-\!  E_d}{ \Gamma(\eta)} + O(\eta^4) \; . \label{eq:Jbulk}
\end{eqnarray}
Furthermore, as we shall be able to verify below, the leading contribution to
the fluctuations of $\eta$ and $\xi$ can be taken independently of
each other (i.e.\ the fluctuations of $\xi$ can be computed assuming
$\eta$ constant, 
and reciprocally).

Subtracting its bulk value from Eq.~(\ref{Consistentxi6}), we
  have  $J(\eta, \xi ) - J^{\rm
  bulk}(\bar \eta, \bar \xi ) \simeq - \bar \eta \delta \eta $, and
thus, by definition of  $\delta J(\eta,\xi)$,
\begin{equation*}
J^{\rm bulk}(\bar \eta + \delta \eta, \bar \xi + \delta \xi) - J^{\rm
  bulk}(\bar \eta, \bar \xi ) = - \delta J(\eta,\xi) - \bar \eta
\delta \eta \; .
\end{equation*}
If the fluctuations of $\xi$ and $\eta$ are small, we can furthermore
approximate $\delta J(\eta,\xi)$ by $\delta J(\bar \eta, \bar \xi)$.
We thus have
\begin{equation} \label{eq:delta_xi}
   \frac{1}{\pi} \frac{\delta \xi}{\Gamma(\bar \eta)} = 
   - \delta J(\bar \eta , \bar   \xi)  +  
  \frac{2}{\pi} \frac{(\bar \xi - E_d)}{\Gamma(\bar \eta)} \frac{\delta
    \eta}{\bar \eta}
- \bar \eta^2 \frac{\delta
    \eta}{\bar \eta}\; .  
\end{equation}
The two last terms on the right-hand-side of Eq.~(\ref{eq:delta_xi})
are proportional to $\bar \eta^2$ [e.g.\ see
Eq.~(\ref{app:deltabarxi}) for the second-to-last term] and so are
negligible in the Kondo regime. Computing the variance $\overline{
  (\delta \xi)^2 }$ therefore amounts, up to the constant factor
$\pi\Gamma(\bar \eta)$, to computing the variance of $\delta J (\bar
\eta , \bar \xi)$.

Now, for $\Gamma(\eta) \gg \Delta$, we have  $u_\kappa = \tilde u_\kappa [\pi\Delta /\Gamma(\eta)] $, where
\begin{equation} \label{eq:uktilde}
\tilde u_\kappa \equiv 
\left[
  \sum_{i=1}^N \frac{x_i
    \Delta^2}{(\lambda_\kappa - \epsilon_i)^2}
\right]^{-1}
\end{equation}
is a dimensionless quantity which for $(\lambda_\kappa\! -\!
\mathcal{E}_0) \ll \Gamma(\eta)$ is essentially independent of $\xi$,
$\Gamma(\eta)$, or the other parameters of the model.  Within the resonance,
and for our random matrix model, we therefore can take the $\tilde
u_\kappa$ to have identical distributions (independent of $\kappa$)
characterized by a variance 
$\sigma^2_u$ of order one.  Neglecting the correlations between the
$\tilde u_\kappa$, and treating the $\kappa$ at the edge of the resonance
as if they were well within it (which is obviously
incorrect but should just affect prefactors that we are in any case
not computing), we have
\begin{equation}
\overline{ (\delta J )^2 } \sim \sum_{-\Gamma < (\lambda_\kappa -
  \mu) < 0} \sigma_u^2 \frac{\pi^2 \Delta^2}{\bar \Gamma^2} \sim (\pi
\sigma_u)^2 \frac{\Delta}{\bar \Gamma} \;  .
\end{equation}
Inserting this into Eq.~(\ref{eq:delta_xi}), we finally get
\begin{equation} \label{eq:xi_scaling}
\overline{ (\delta \xi)^2 } \sim \bar\Gamma \Delta \sim \Tk \Delta \; .
\end{equation}
With regard to the limits of validity of this estimate,
note that our random matrix model (Sec.\,\ref{sec:RMT})
assumes implicitly that the Thouless energy $E_{\rm Th}$ is infinite, and more
specifically that $E_{\rm Th} \gg T^{\rm bulk}_K$. For a
chaotic ballistic system with $E_{\rm Th} \ll T^{\rm bulk}_K$,
the $\tilde u_\kappa$ are independent only in an
interval of size $E_{\rm Th}$; thus, Eq.~(\ref{eq:xi_scaling}) should be
replaced by $\overline{ (\delta \xi)^2 } \sim E_{\rm Th}  \Delta$.

For the fluctuations of $\eta$, we proceed in a similar way,
subtracting Eq.~(\ref{Consistenteta6}) from its bulk analog and
assuming small fluctuations, and so find
\begin{equation} \label{eq:eta_fluct1}
  \frac{\delta \eta}{\bar \eta} = \frac{\delta \Gamma}{2 \bar
    \Gamma}
= \frac{1}{4} \left[ \delta I(\bar \eta, \bar \xi) - \frac{\pi\delta
    \xi}{\Gamma_0} \right]\; .  
\end{equation}
Here, however, it is necessary to split the sum over states in
Eq.~(\ref{Consistenteta6}) into two parts: $I = I^{\rm in} + I^{\rm out}$
where $I^{\rm in}$ and $I^{\rm out}$ are defined in the same way as
$I$ but over an energy range corresponding, respectively, to the inside
and outside of the resonance.  One has 
$I^{\rm out}(\eta,\xi) \gg I^{\rm in}(\eta,\xi)$  since the former
contains the logarithmic divergence.  However, the fluctuations of the
two quantities are of the same order [basically because when considering
the variance, and thus squared quantities, one  transforms a diverging sum
$\sum_\kappa (\lambda_\kappa - \mathcal{E}_0)^{-1}$ into a converging
one $\sum_\kappa (\lambda_\kappa - \mathcal{E}_0)^{-2}$].
Indeed, the sum $I^{\rm out}(\eta,\xi)$ is, up to sub-leading
corrections, the same as the one  entering into the definition of
$\Tk$.  Its fluctuations have been evaluated in
Refs.\, \onlinecite{KaulEPL05}-\onlinecite{Kettemann04}, leading to 
\begin{equation}
 \overline{ (\delta I^{\rm out})^2 } \sim
 \frac{\Delta}{\Tk^{\rm bulk}} \; ,
\end{equation}
which is consistent with Eq.~(\ref{eq:Tk_fluct}).
The variance of $\delta I^{\rm in}$ can, on the other hand, be evaluated
following the same route as for $\delta J$, yielding
\begin{equation}
\overline{ (\delta I^{\rm in} )^2 } \sim 
2\sum_{ 0  < (\lambda_\kappa -  \mu) < \Gamma} \frac{\sigma_u^2}{(\bar
  \Gamma/\pi \Delta)^{2}} \frac{\pi^2 }{\Gamma(\eta)^2}  (\lambda_\kappa -
\mathcal{E}_0)^2 
\sim  \frac{\Delta}{\bar\Gamma} \;  . 
\end{equation} 
This shows, then, that the two contributions $\overline{ (\delta
  I^{\rm in})^2}$ and $\overline{ (\delta I^{\rm out})^2}$ scale in
the same way. 

For the final contribution---the last term on the r.h.s.\ of Eq.~(\ref{eq:eta_fluct1})---Eq.~(\ref{eq:xi_scaling}) implies
\begin{equation}
 \frac{\pi\overline{(\delta \xi )^2 }}{\Gamma_0} \sim 
 \eta^2 \frac{\Delta}{\bar\Gamma} \;  ,
\end{equation}
which is proportional to ${\Delta}/{\bar\Gamma}$ as for the first two
contributions, but the extra smallness factor $\eta^2$ makes it
negligible in the Kondo limit. Gathering everything together, we
therefore obtain
\begin{equation} \label{eq:eta_fluct2}
  \frac{\overline{(\delta \eta)^2}}{\bar \eta^2} = 
 \frac{\langle(\delta \Gamma)^2\rangle }{4 \bar
    \Gamma^2}
    \sim \frac{\Delta}{\bar\Gamma} \; .
\end{equation}
[If $E_{\rm Th} \ll \Gamma(\eta)$, $\overline{(\delta \xi )^2 }$ and 
$\overline{ (\delta I^{\rm in} )^2 } $ are reduced by a factor
$(E_{\rm Th}/\bar \Gamma)$, but not $\overline{ (\delta I^{\rm out} )^2
}$; thus, Eq.~(\ref{eq:eta_fluct2}) remains unchanged.] 

Turning to the mixed-valence regime by releasing the constraint $\eta
\ll 1$, we see that $\pi\overline{(\delta \xi )^2 }/\Gamma_0$ becomes
comparable in size to the other contributions to $\overline{(\delta
  \eta )^2}$ and has the same parametric dependence.  Furthermore,
taking the derivative $\partial J^{\rm bulk} /
\partial \xi$ [see Eq.~(\ref{app:Jbulk})] implies that the left hand
side of (\ref{eq:delta_xi}) should be multiplied by a factor $\bar \Gamma /
(\bar \Gamma^2 - (\bar \xi - \mathcal{E}_0)^2)$, which, however, does not
change the scaling of $\overline{(\delta \eta )^2}$.  In the same
way, using Eq.~(\ref{eq:eta_fluct2}) the two last terms 
on the right hand side of Eq.~(\ref{eq:delta_xi}), which are
proportional to  $\delta \eta/\bar \eta$, give a contribution
$\sim \eta^4 \bar \Gamma \Delta$ to $\overline{(\delta \xi )^2 }$,
as well as the term $(\bar \xi - E_d) \delta \xi / (\bar \Gamma^2 -
(\bar \xi - \mathcal{E}_0)^2)$ that should be added to the the left
hand side of Eq.~(\ref{eq:eta_fluct1}) from   $\partial I^{\rm bulk} /
\partial \xi$ [see Eq.~(\ref{app:Ibulk})]. Those are negligible in the
Kondo regime, but are of the same size and with the same scaling as
the contribution due to $\delta J$ in the mixed-valence regime.  We
find, then, that the fluctuations of the mean field parameters scale
with system size in the same way in both the Kondo and mixed-valence
regimes: the variance of both $\xi$ and $\eta$ is proportional to
$\Delta$.

\subsection{Numerical investigations}

\begin{figure}[t]
\centering
\includegraphics[width=3.2in,clip]{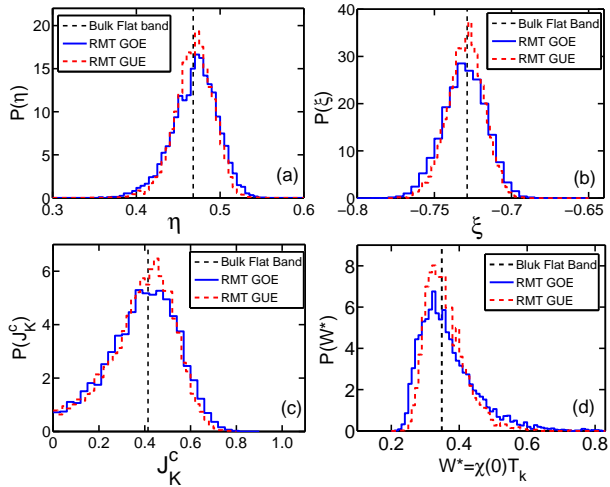}
\caption{(Color online) The distribution of the mean field parameters,
  critical coupling, and effective Wilson number from the SBMFT
  calculation for both GOE (blue solid line) and GUE (red dash-dotted
  line). (a) $\eta$, (b) $\xi$, (c) critical coupling $J_K^c$, and (d)
  the effective Wilson number $W^{*}=\Tk\chi_0(0)$.  The vertical
  black dashed lines mark the values for the corresponding bulk
  flat-band system. The following parameters were used: band width
  $D=3$, $E_d=-0.7$, $V_0=0.6$, and $T=0.005$. The mean level
  spacing is $\Delta=0.01$, and the Kondo temperature in the bulk
  limit is $\Tk^{\rm bulk}\simeq 0.092$.  
}
\label{fig:MF_para}
\end{figure}

\begin{figure}[t]
\centering
\includegraphics[width=3.2in,clip]{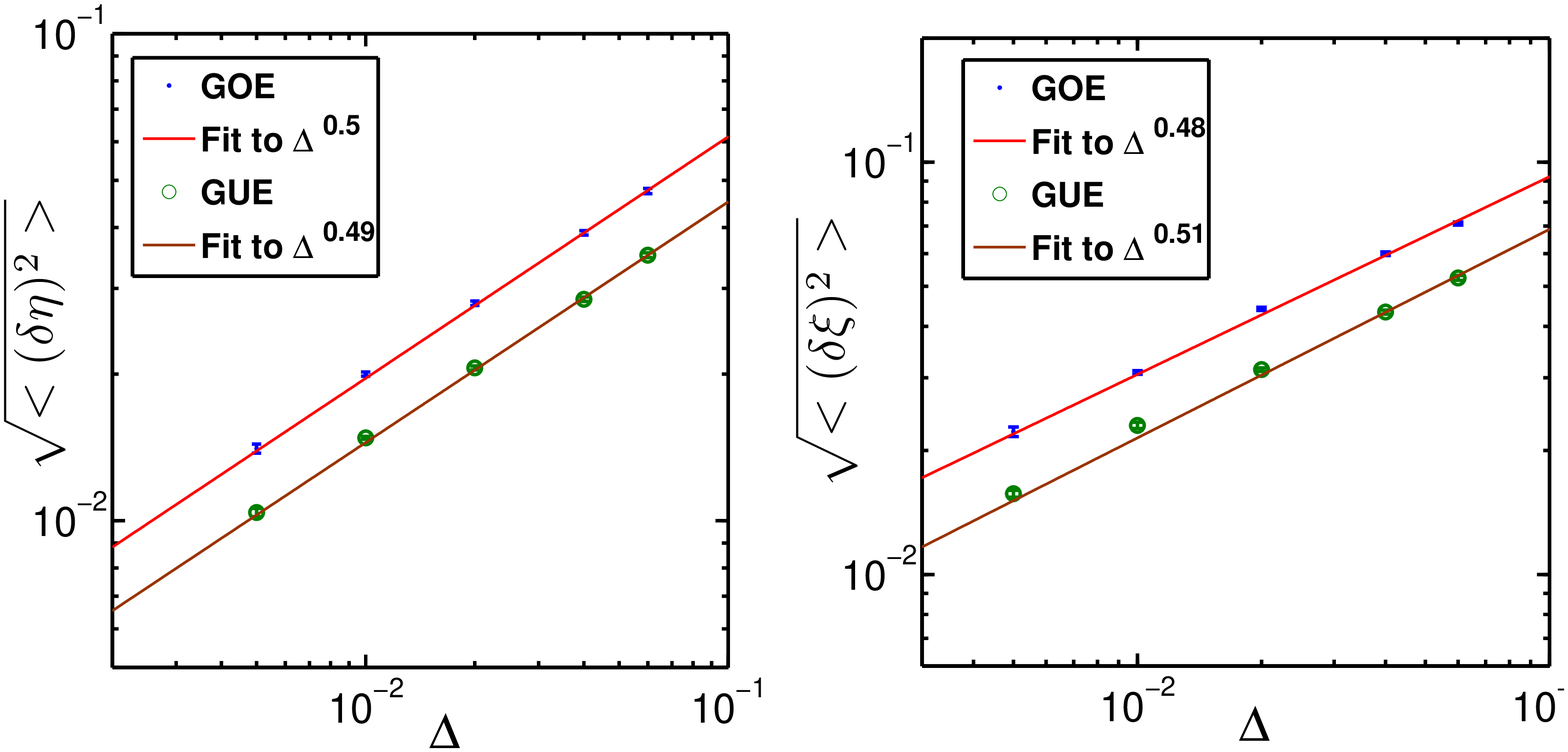}
\caption{(Color online) Variance of the mean field parameters $\eta$
  (left panel) and $\xi$ (right) as the size of the system is
  changed. The change in system size is quantified through the mean
  level separation $\Delta$. As expected, the fluctuations are smaller
  in the GUE compared to the GOE, and the dependence of the variance
  of the fluctuations on $\Delta$ is nearly linear. Here, we use
  $D=3$, $E_d=-0.7$, and $V_0=0.8$. }
\label{fig:MF_scaling}
\end{figure}

To illustrate the previous discussion, we have computed numerically
the self-consistent parameters $\eta$ and $\xi$ for a large number of
realizations of our random matrix ensemble at various values of
the parameters defining the Anderson box model 
(always within our regime of interest, $T\! < \! \Delta \!
\ll \! \Tk$, except when explicitly specified).  
Fig.~\ref{fig:MF_para} shows the distributions of
$\eta$ and $\xi$ for a choice of parameters such that $\Tk^{\rm bulk} / \Gamma_0 \simeq 0.24$ (close to but not in the mixed valence
regime).  We see that these distributions are approximately Gaussian
and centered on their values for the bulk flat-band case, though note
the slightly non-Gaussian tail on the left side in both cases. The
distributions for the GOE and GUE are qualitatively similar, with
those for the GUE being, as expected, slightly narrower. As
anticipated, the fluctuation of these mean parameters is small: the
root-mean-square variation is less than 5\% of the mean.
Fig.~\ref{fig:MF_scaling} further shows how the variance of $\eta$ and
$\xi$ varies with the parameters of the model, confirming the behavior in
Eqs.~(\ref{eq:xi_scaling}) and (\ref{eq:eta_fluct2}).

%%%%%%%%%%%%%%%%%%%%%%%%%%%%%%%%%%%%%%%%%%%%%%%%%%%%%%%%%%%%%%%%%%%%%%%%%%%%
\section{Other global physical properties}
\label{sec:globalphys}

Beyond $\eta$ and $\xi$ themselves, several interesting global
properties of the system follow directly from the solution of the
mean-field problem. We briefly discuss two of them here.

\subsection{Wilson number:  Comparing $\Tk$ and the ground state
  properties} 

The ``Wilson number'' is an important quantity in Kondo
physics: it compares $\Tk$ with the energy scale contained in the
ground state magnetic susceptibility. It is defined as $W^{*} \equiv
\Tk\, \chi_0({ T\!\rightarrow \!  0})$, where $\chi_{0}(T)\equiv
\int_{0}^{1/T} \langle 
S^{z}(\tau)S^{z}(0)\rangle d\tau$ is the static susceptibility.  $W^{*}$
is thus the ratio between the characteristic high temperature scale
$\Tk$ and the characteristic low temperature scale $T_0 =
1/\chi_0(T=0)$ of the strong-coupling regime \cite{BurdinRev2009}.

In the bulk Kondo problem, there is only one scale, of course,
and so the Wilson number has a fixed value \cite{HewsonBook}, namely
$0.4128$ (approximated as $0.349$ in the SBMFT).  
For our mesoscopic Anderson box on the other hand, this
will be a fluctuating quantity that has to be computed for each
realization of the mesoscopic electron bath.  Computing $\Tk$
according to Eq.~(\ref{eq:TKAndersonbox2}) and expressing the static
susceptibility as $\chi_{0}(T)= T \sum_{n=-\infty}^{+\infty}
G_{ff}(i\omega_{n})G_{ff}(-i\omega_{n})$ with $G_{ff}(i\omega_{n})$ given by
Eq.~(\ref{AndersonboxGff1}), we
obtain the distribution of the Wilson number shown in
Fig.~\ref{fig:MF_para}(d).  Note the unusual non-Gaussian form
of the distribution, with the long tail for large $W^*$. As a result,
the peak of the distribution is slightly smaller than the bulk
flat-band value. The magnitude of the fluctuations in $W^*$ is modest
for our choice of parameters (about 30\%) but considerably larger than
the magnitude of the fluctuations of the mean field parameters in
Fig.~\ref{fig:MF_para}(a)-(b).

\subsection{Critical Kondo coupling}
\label{sec:JKc}

Another interesting global quantity is the critical
Kondo coupling $J_{K}^{c}[\{\epsilon_i\},\{|\phi_i(0)|^2\}]$ defined for
a given realization of the electron bath by
\begin{equation} \label{eq:JKc}
 \frac{1}{2 J_{K}^{c}} \equiv
\sum_{i=1}^{N}
\frac{\vert \phi_{i}(0)\vert^{2}}{|\epsilon_{i}-\mu|} \; .
\end{equation}
Here, exceptionally, we move away from the regime 
$\Tk^{\rm bulk} \gg \Delta$.
The discreetness of the spectrum is what is making convergent the sum in the
above expression, and thus $J_{K}^{c}$ can be defined only
because of the finite size of the electron bath.

Comparing with Eq.~(\ref{eq:TKAndersonbox2}), we see that in the SBMFT
approximation ${J_{K}^{c}}$ is the realization-dependent value of
the Kondo coupling $J_K$ [defined in Eq.~(\ref{eq:JK})] such that $\Tk=0$ 
if $J_K < J_{K}^{c}$ and $\Tk$ is non-zero if $J_K > J_{K}^{c}$.
Note that the possibility of vanishing the Kondo temperature $T_K$
  has been discussed in the framework of disordered bulk
  systems \cite{Kettemann04,Kettemann06,Kettemann07,Zhuravlev08, Kondodisorder1, Kondodisorder2}. 
Fig.~\ref{fig:MF_para}(c) shows the distribution of this critical
coupling for a mesoscopic Anderson box. Note the non-Gaussian form of
the distribution and the similarity between the GOE and
GUE results. Remarkably the distribution functions do not vanish at
$J_{K}^{c}=0$, indicating that there exist realizations for
which the Kondo screening occurs for any coupling $V_0$ and impurity
level $E_d$. Indeed, as pointed out in Ref.\,\onlinecite{Bedrich10},
a small $J_{K}^{c}$ corresponds to a situation in which the chemical
potential $\mu$ lies very close to some level $\epsilon_i$, which then
dominates the sum in (\ref{eq:JKc}).  If $\mu$ exactly coincides with
some $\epsilon_i$, $J_{K}^{c}=0$: the large dot contains an odd number
of electrons on average so the impurity can always form a singlet
with the large dot \cite{KaulPRL06}.

%%%%%%%%%%%%%%%%%%%%%%%%%%%%%%%%%%%%%%%%%%%%%%%%%%%%%%%%%%%%%%%%%%%%%%%%%%%%
\section{Spectral fluctuations}
\label{sec:spectra}

The mean field approach maps the Kondo problem at low temperatures into
a resonant level problem, Eq.~(\ref{MFHamiltonianAndersonbox1}), with
two realization specific parameters: the energy of the resonant level
[$\mathcal{E}_0(\xi)$, taking $\mu=0$ as the energy reference] and the
strength of the coupling to it. We have seen, however, that in the
limit $\Tk \gg \Delta$ [or equivalently $\Gamma(\eta) \gg \Delta$] the
scale of the fluctuations of these parameters both go to zero as
$\sqrt{\Delta/\bar\Gamma}$.  Furthermore, as long as $|\lambda_\kappa
-\mathcal{E}_0| \ll \Gamma(\eta)$, the $\lambda_\kappa$ and
corresponding $|\psi_\kappa\rangle$ are relatively insensitive to
$\Gamma$ and $\mathcal{E}_0$ and thus to their fluctuations. We
consider, therefore, in a first stage the fluctuations implied by the
resonant level model (RLM) with \emph{fixed} parameters, and then come
back later to consider how the fluctuations of the parameters modify
the results.

For the analysis in this section and the next, it is convenient to
rewrite the resonant level model (RLM) as 
\begin{equation}
\label{eq:RLM}
H_{\rm RLM} \!=\! \sum_{i=1}^N \epsilon_i |i\rangle\langle i|
        + \epsilon_0|f\rangle\langle f|
      + v \sum_{i=1}^N [\sqrt{N}\phi_i(0)|i\rangle\langle f|+ {\rm h.c.}] .
\end{equation}
Here, $|f\rangle$ is the bare resonant level state with energy
$\epsilon_0$, and the $|i\rangle$ for $i \ge 1$ are the bare
(unperturbed) states of the reservoir with wave functions $\phi_i
({\bf r})$. The eigenstates of $H_{\rm RLM}$ (perturbed states) are,
as before, $|\psi_\kappa\rangle$ for $\kappa=0, \cdots ,N$ with
corresponding eigenvalues $\{\lambda_\kappa\}$. Finally, the coupling
strength is taken to scale with system size as $v \propto 1/\sqrt{N}$
so the large $N$ limit in the random matrix model can be
conveniently taken.  The corresponding width of the resonant level is
$\Gamma \equiv \pi \rho_0 N v^2$.

We use two complementary ways of viewing the RLM. First, as a
microscopic model in its own right, albeit non-interacting, one has
$v=V_0/\sqrt{N}$ where $V_0$ is the hopping matrix element from the
resonant level to the ${\bf r}=0$ site in the reservoir as in
Eq.~(\ref{HimpAndersonboxU1}). In this case the width of the level is
simply $\Gamma = \Gamma_0$, and $\epsilon_0$ is just a parameter of
the model. Second, if one views the RLM as the result of an SBMFT
approach in which the fluctuations of the mean field parameters are
neglected, one has $v= \bar\eta V_0/\sqrt{N}$, in which case $\Gamma =
\Gamma(\bar\eta) = \bar\Gamma$, and $\epsilon_0 = \mathcal{E}_0(\bar
\xi)$.  We stress that in both views, $\epsilon_0$ and the
$\epsilon_i$'s ($1 \leq i \leq N$) are, in spite of the similarity in
the notations, different objects in terms of the statistical ensemble
considered: $\epsilon_0$ is a fixed parameter, when the $\epsilon_i$'s
are random variables distributed according to Eq.~(\ref{eq:WD_dist}).

\subsection{Joint Distribution Function}

To characterize the correlations between the unperturbed energy levels
and the perturbed levels, the basic quantity needed is the joint
distribution function $P(\{\epsilon_i\},\{\lambda_\kappa\})$. As seen
in Sec.\,\ref{sec:preliminary}, the RLM eigenvalues $\lambda_\kappa$
are related to the unperturbed energies through
Eq.~(\ref{eq:lambdas}), which we rewrite as
\begin{equation}
\label{eq:EL_relation}
     \sum_{i=1}^{N}\frac{x_i}{\lambda_\kappa-\epsilon_i} 
          = \frac{\lambda_\kappa-\epsilon_0}{v^2}, 
\end{equation}
remembering that $x_i \equiv N |\phi_i(0)|^2$.  Explicitly writing out
the ``interleaving'' constraints, we obtain
 \begin{eqnarray}
 \label{eq:EL_condition}
 \epsilon_i \leq & \lambda_i & \leq \epsilon_{i+1} , \qquad
 i=1,\cdots,N-1 \nonumber\\ 
 & \lambda_0 & < \epsilon_1 \\
 & \lambda_N & > \epsilon_N  \; . \nonumber
 \end{eqnarray}
 (Note we slightly change the way we index the levels $\lambda_i$ with
 respect to section~\ref{sec:MF}.)  There is furthermore an additional
 constraint on the sum of the eigenvalues
\begin{equation}
\label{eq:EL_sum_condition}
\mathcal{D} \equiv \sum_{i=0}^N \epsilon_i - \sum_{\kappa=0}^N
\lambda_\kappa =0\; ,
\end{equation}
a proof of which is given in Appendix~\ref{app:EVconstraint}.

Since we know the joint distribution of the $\epsilon_i$ and
$|\phi_i(0)|^2$, we now want to use relation (\ref{eq:EL_relation}) to
convert from the eigenfunctions to the $\lambda_\kappa$. A slight
complication here is that there is one more level $\lambda_\kappa$
than wavefunction probabilities $|\phi_i(0)|^2$ (which is why a
constraint such as Eq.~(\ref{eq:EL_sum_condition}) needs to
appear). It is therefore convenient to include an additional
``unperturbed'' level at energy $\epsilon_0 $ associated with a
  wave-function probability $x_0$, and to extend the summation in the
  left hand side of Eq.~(\ref{eq:EL_relation}) to $i=0$.  Assuming
  then that $x_0$ has a probability one to be zero [i.e. that $P(x_0)
  = \delta(x_0)$], one recovers the original problem.

In terms of the Jacobian for this variable transformation, the desired joint
distribution then can be written as
\begin{equation}
\label{eq:joint_ini}
P_\beta(\{\epsilon_i\},\{\lambda_\kappa\}) = P_\beta(\{\epsilon_i\})
\; \delta(x_0) \;
\prod_{i=1}^N p_\beta(x_i)  \; \bigg\arrowvert \det\left[\frac{\partial
    x_i}{\partial\lambda_j}\right] \bigg\arrowvert 
\end{equation}
where $ P_\beta(\{\epsilon_i\})$ and $p_\beta(x_i)$ are given in
Eqs.~(\ref{eq:WD_dist}) and (\ref{eq:PT}). (We shall not assume
in this subsection that the spectrum $\{\epsilon_i\}$ has been
unfolded.) In order to find the Jacobian, we first find $x_i$
explicitly. Since Eq.~(\ref{eq:EL_relation}) is linear in $x_i$,
inverting the Cauchy matrix $a_{\kappa i}=
1/(\lambda_\kappa-\epsilon_i)$ yields
\begin{equation}
x_i=\sum_{\kappa} b_{i\kappa}\frac{\lambda_\kappa-\epsilon_0}{v^2}~, 
\end{equation}
where \cite{Schechter59}
\begin{eqnarray}
 b_{i\kappa} &=&
 \frac{1}{\epsilon_i-\lambda_\kappa}
\frac{A(\epsilon_i)}{B^{'}(\epsilon_i)}
\frac{B(\lambda_\kappa)}{A^{'}(\lambda_\kappa)}~,  \nonumber\\ 
A(z) &=& \prod_{\kappa=0}^N(z-\lambda_\kappa), \qquad B(z) =
\prod_{i=0}^N(z-\epsilon_i)  \; .
\end{eqnarray}
This expression can be simplified by using the residue theorem twice. First,
note that 
\begin{eqnarray}
 x_i 
&=& -\frac{1}{v^2} \bigg( \frac{1}{2\pi i}\oint\frac{(z-\epsilon_0)
  B(z)}{(z-\epsilon_i) A(z)}dz \bigg)
\frac{A(\epsilon_i)}{B^{'}(\epsilon_i)} \;. 
\end{eqnarray}
Second, the identity $\oint\prod_i(z-a_i)/\prod_i(z-b_i)dz=2\pi i
\sum_i(b_i-a_i)$ implies
\begin{equation}
\label{eq:x_final}
 x_i=\frac{1}{v^2} \left(\epsilon_i-\epsilon_0 + \mathcal{D}
        \right)
 \frac{\prod_{\kappa=0}^N(\lambda_\kappa-\epsilon_i)}
      {\prod_{i\neq j}(\epsilon_j-\epsilon_i)} \; .
\end{equation}
For $i=0$, this reads
\begin{equation}
 x_0=\frac{1}{v^2}  \mathcal{D} \cdot        
 \frac{\prod_{\kappa=0}^N(\lambda_\kappa - \epsilon_0)}
      {\prod_{j=1}^N (\epsilon_j - \epsilon_0)} \; , 
\end{equation}
and thus $x_0 \neq 0$ implies that the $\lambda_\kappa$ cannot coincide with
$\epsilon_0$, leading then to 
\begin{equation}
\label{eq:deltax0}
\delta (x_0) =  \frac{v^2 \prod_{j=1}^N(\epsilon_j-\epsilon_0)}
{\prod_{\kappa=0}^N (\lambda_\kappa- \epsilon_0)} \cdot 
 \delta(\mathcal{D}) 
      \;  . 
\end{equation}
The factor $\delta (x_0)$ in Eq.~(\ref{eq:joint_ini}) therefore imposes the constraint
(\ref{eq:EL_sum_condition}) that we know should hold.

Now note that $\partial x_i / \partial\lambda_\kappa$ is itself a
Cauchy-like matrix $\partial x_i / \partial\lambda_\kappa = r_i s_\kappa /
(\lambda_\kappa-\epsilon_i)$ where 
\begin{equation}
 r_i=\frac{1}{v^2}\frac{\prod_\kappa(\lambda_\kappa-\epsilon_i)}{\prod_{\kappa
     \neq i}(\epsilon_\kappa-\epsilon_i)} \quad\text{and}\quad
 s_\kappa=\lambda_\kappa - \epsilon_0 + \mathcal{D} \;.
\end{equation}
The Jacobian, then, is given by
\begin{eqnarray}
 \label{eq:jacobian}
 \det\bigg(\frac{\partial x_i}{\partial\lambda_\kappa} \bigg) &=& 
  \prod_{i=0}^{N} r_i \prod_{\kappa=0}^{N} s_\kappa
 \det\bigg( \frac{1}{\lambda_\kappa-\epsilon_i} \bigg)\nonumber\\
&=&
\frac{\prod_\kappa(\lambda_\kappa-\epsilon_0 + \mathcal{D} )}{v^{2N}}
\frac{\prod_{j>i}(\lambda_j-\lambda_i)}{\prod_{j>i}(\epsilon_j-\epsilon_i)}
\;. 
\end{eqnarray}

From now on, since no further derivative will be taken, we can set
$x_0$, and thus $\mathcal{D}$, to zero, and thus assume that the constraint
(\ref{eq:EL_sum_condition}) holds.  The last ingredient we need in
order to assemble the joint distribution function is $\sum_i x_i$:
\begin{eqnarray}
 \label{eq:x_sum_ini}
\sum_i x_i &=& \frac{1}{v^2}\sum_i
(\epsilon_0-\epsilon_i)\frac{A(\epsilon_i)}{B^{'}(\epsilon_i)}\nonumber\\ 
&=& -\frac{1}{2\pi i v^2}\oint\frac{(z-\epsilon_0)A(z)}{B(z)}dz\nonumber\\
&=& -\frac{1}{2\pi i
  v^2}\oint\frac{\prod_{\kappa=0}^N(z-\lambda_\kappa)}
{\prod_{i=1}^N(z-\epsilon_i)}dz 
\;.  
\end{eqnarray}
The relation
\begin{eqnarray}
\lefteqn{ \frac{1}{2\pi i}
  \oint\frac{\prod_{i=1}^N(z-a_i)}{\prod_{i=1}^{N-1}(z-b_i)}dz } &&
\nonumber\\ 
 &&  = \frac{1}{2}\bigg[\displaystyle\sum_{i=1}^{N-1}
 b_i^2-\displaystyle\sum_{i=1}^{N}a_i^2 + 
\Big(\displaystyle\sum_{i=1}^N a_i - \displaystyle\sum_{i=1}^{N-1} b_i
\Big)^2 \,\bigg] 
\end{eqnarray}
and the sum constraint (\ref{eq:EL_sum_condition}) then gives
\begin{equation}
 \label{eq:sum_x_final}
\sum_i x_i = -\frac{1}{2v^2}
\bigg (\sum_{i=0}^N \epsilon_i^2-\sum_{\kappa=0}^N \lambda_\kappa^2 \bigg) \;.
\end{equation}
Finally, assembling all the different elements, Eqs.\ (\ref{eq:WD_dist}), 
(\ref{eq:PT}), (\ref{eq:x_final}), (\ref{eq:jacobian}), and
(\ref{eq:sum_x_final}), we arrive at the desired result for the joint
distribution function: within the domain specified in
(\ref{eq:EL_condition}), 
\begin{widetext}
\begin{equation}
 \label{eq:joint_final}
P_\beta(\{\epsilon_i\},\{\lambda_\kappa\}) \propto  
\frac{ \displaystyle \prod_{i>j \ge 1}(\epsilon_i-\epsilon_j)
  \prod_{\kappa>\nu \ge 0}(\lambda_\kappa-\lambda_\nu)}
{ \displaystyle \prod_{i=1}^N \prod_{\kappa=0}^N
  |\epsilon_i-\lambda_\kappa|^{1-\beta/2}}  \,
\delta\Big(\sum_{\kappa=0}^N \lambda_\kappa- \sum_{i=0}^N
  \epsilon_i \Big)
 \exp\bigg[-\frac{\beta}{4v^2} \left(\sum_{\kappa=0}^N \lambda_\kappa^2
 - \sum_{i=0}^N \epsilon_i^2 \right) \bigg]
\exp\bigg[-\frac{1}{4\alpha^2}   \sum_{i=1}^N \epsilon_i^2 \bigg]
   \;.
\end{equation}
\end{widetext}
(In the last exponential, $\alpha = \sqrt{N}\Delta/\pi$.)
We stress again that in Eq.~(\ref{eq:joint_final}), $\epsilon_0$ is
not a random variable, but a fixed parameter.

\subsection{Toy models}

The joint distribution Eq.~(\ref{eq:joint_final}) contains in
principle all the information about the spectral correlations between
the high and low temperature spectra of the mesoscopic Kondo problem.
It is, however, not straight forward here, as in other circumstances
(cf.\ Ref.\,\onlinecite{MehtaBook}), to deduce from it explicit
expressions for basic correlation properties.  Instead of pursuing
this route, we shall here follow the spirit of the Wigner approach to
the nearest neighbor distribution of classic random matrix ensembles
\cite{Bohigas91} and introduce a simple toy model, easily solvable,
which provides nevertheless good insight for some of the correlations
in the original model.

Starting from Eq.~(\ref{eq:EL_relation}) for the level
$\lambda_\kappa$ of the RLM, we first notice that the resonance width
$\Gamma = \pi \rho_0 N v^2$ defines two limiting
regimes. When $\lambda_\kappa$ is well outside the resonance,
$|\lambda_\kappa - \epsilon_0 | \gg \Gamma$, the low temperature
level $\lambda_i$ has to be (almost) equal to $\epsilon_i$ or
$\epsilon_{i+1}$; as expected, the two spectra nearly coincide. On the
other hand, well within the resonance, $|\lambda_\kappa -
\epsilon_0| \ll \Gamma$ so the r.h.s.\ of
(\ref{eq:EL_relation}) can be set equal to zero,
\begin{equation}
\label{eq:Connection_toy1}
 \sum_{i=1}^N\frac{x_i}{\lambda_\kappa - \epsilon_i} \approx 0 \;,
\end{equation}
thus providing a first simplification.  

Let us now consider the level $\lambda_\kappa$ located between
$\epsilon_i$ and $\epsilon_{i+1}$. It is reasonable to  
assume that the position of $\lambda_\kappa$ will be mainly determined
by these two levels and the fluctuations of their  
corresponding eigenfunctions $|\phi_{i}(0)|^2 =  x_i/N$ and
$|\phi_{i+1}(0)|^2 = x_{i+1}/N$, and
that the influence of the other states will be significantly weaker.  
Neglecting completely the influence of all but these closest
$\epsilon$'s, the problem then reduces to the much simpler equation for
 $\lambda_\kappa$,
\begin{equation} \label{eq:Connection_toy_bis}
   \frac{x_i}{\lambda_\kappa - \epsilon_i}
+  \frac{x_{i+1}}{\lambda_\kappa - \epsilon_{i+1}}  = 0 \; ,
\end{equation}
where $x_i$ and $x_{i +1}$ are uncorrelated and distributed according
to the Porter-Thomas distribution (\ref{eq:PT}). One notices then that
all energy scales ($v$, $\Delta$, etc. ...) have disappeared from the
problem except for $\epsilon_{i+1} -\epsilon_i$.  The resulting
distribution of $\lambda_\kappa$ is therefore universal, depending
only on the symmetry under time reversal. Straightforward integration
over the Porter-Thomas distributions gives
\begin{eqnarray}
P(\lambda_\kappa) &  = &  \frac{1}{\pi} 
\frac{1}{\sqrt{(\epsilon_{i+1} - \lambda_\kappa)(\lambda_\kappa
    - \epsilon_i)}} \qquad \mbox{GOE} \label{eq:PtoyGOE}\\
P(\lambda_\kappa) &  = &  \frac{1}{\epsilon_{i+1} -\epsilon_i} 
\qquad \qquad \mbox{GUE} \; . \label{eq:PtoyGUE}
\end{eqnarray}
Breaking time-reversal invariance symmetry thus affects drastically the
correlation between  the low temperatures level $\lambda_\kappa$ and
the neighboring high temperatures ones $\epsilon_i$ and
$\epsilon_{i+1}$. Time-reversal 
symmetric systems see a clustering of
the $\lambda_\kappa$'s close to the $\epsilon_i$'s---with a square root singularity---while for systems
without time-reversal symmetry the distribution is uniform between 
$\epsilon_i$ and $\epsilon_{i+1 }$.

In the GUE case, for which the Porter-Thomas distribution is
particularly simple, we can consider a slightly more elaborate version
of our toy model.  It is, for instance, possible to include the average effect of
all levels beyond the two neighboring ones (for which we keep the
fluctuations of only the wave-functions, not the energy levels). 
Furthermore one can take into account the term $(\lambda_\kappa -
\epsilon_0)/v$ that was neglected above, assuming that its
variation in the interval $[\epsilon_i,\epsilon_{i+1}]$ is small.  Introducing
$\bar \lambda \equiv(\epsilon_i\!+\!\epsilon_{i+1})/2$ and $\sigma
\equiv (\lambda_\kappa \!- \!\bar \lambda)/\Delta \in [-1/2, +1/2]$,
Eq.~(\ref{eq:Connection_toy_bis}) is replaced by
\begin{equation} \label{eq:toy2}
   \frac{x_i}{\sigma +\frac{1}{2}}
+  \frac{x_{i+1}}{\sigma -\frac{1}{2}}   =  \mathcal{F}(\sigma)~, 
\end{equation}
with
\begin{eqnarray}
 \mathcal{F}(\sigma)   & \equiv  & 
\sum_{i \neq 0,1} \frac{1}{\sigma+\frac{1}{2}-i} + \pi \frac{\bar
  \lambda}{\Gamma} \\
& = & \pi \left( \tan(\pi \sigma) + \frac{\bar \lambda}{\Gamma}
\right) 
- \left(\frac{1}{\sigma +\frac{1}{2}}
+  \frac{1}{\sigma -\frac{1}{2}} \right) \;. \nonumber
\end{eqnarray}
Integrating over the Porter-Thomas distribution, we obtain in the GUE
case
\begin{eqnarray} \label{eq:Ptoy2}
P(\sigma)   &=&   \exp\left[ -\big(\sigma+\frac{1}{2}\big) \mathcal{F}(\sigma)-u_{\rm min}\right] 
   \\
&\times& \left[1 + u_{\rm min} +  (\frac{1}{2}+\sigma) \mathcal{F}(\sigma)
+ (\frac{1}{4} + \sigma^2) \frac{d \mathcal{F}}{d \sigma} \right] \nonumber 
\end{eqnarray}
with $u_{\rm min} \equiv \inf[0,-\mathcal{F}(\sigma)]$. Replacing
$\mathcal{F}(\sigma)$ by zero in Eq.~(\ref{eq:Ptoy2}) of course
recovers  Eq.~(\ref{eq:PtoyGUE}).

\subsection{Numerical distributions}

\begin{figure}[t]
\centering
\includegraphics[width=3.4in,clip]{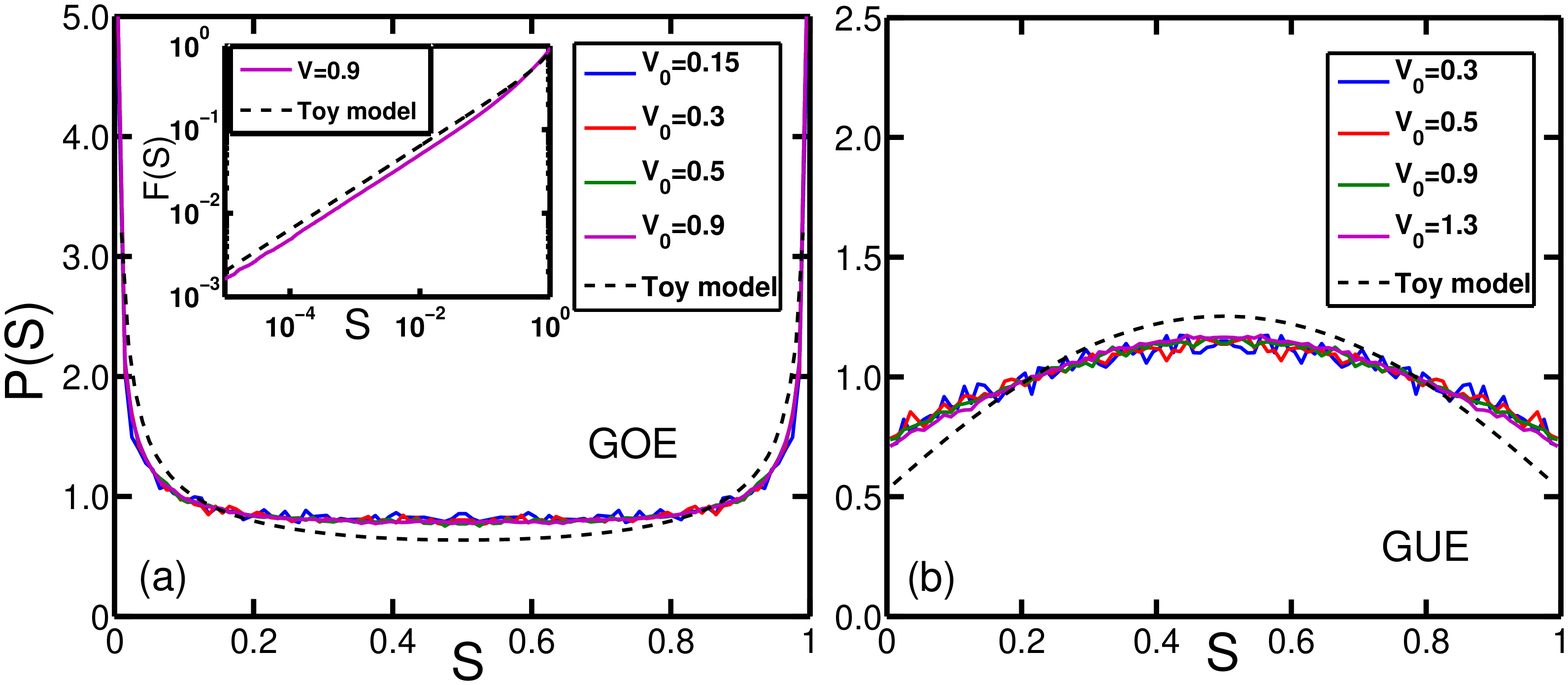}
\includegraphics[width=3.4in,clip]{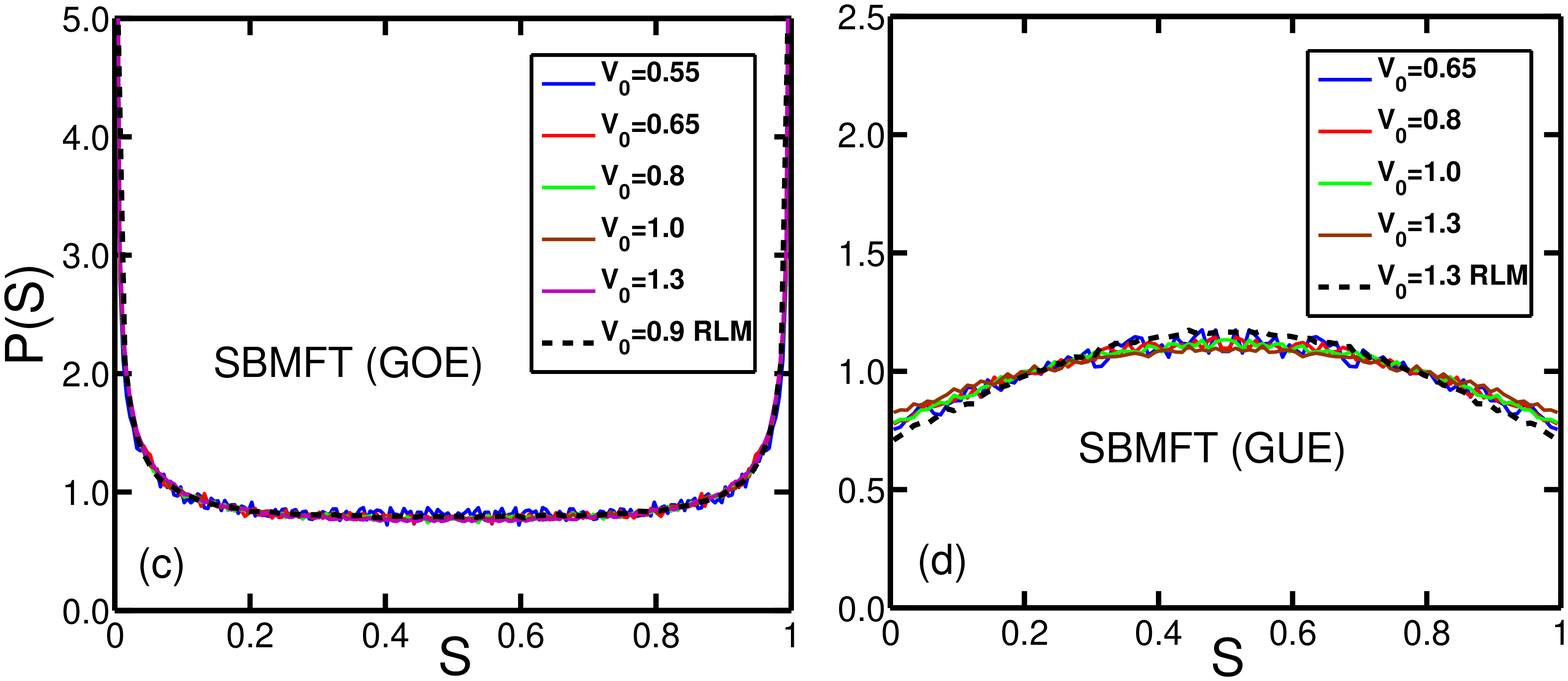}
\vspace*{-0.2in}
\caption{(Color online) The distribution of $S$ (which includes
    both $|\lambda_\kappa-\epsilon_i|/|\epsilon_{i+1}-\epsilon_i|$ and
    $|\lambda_\kappa-\epsilon_{i+1}|/|\epsilon_{i+1}-\epsilon_i|$) for
    the resonant level model (a) GOE, (b) GUE, and for the SBMFT
    treatment of the infinite-$U$ Anderson model (c) GOE, (d) GUE.
    Insert of (a): the cumulative distribution of the $V_0=0.9$ GOE
    data compared to the toy model. The dashed lines in (a) and (b)
    are the result of the toy model; those in (c) and (d) show the RLM
    result for $V_0=0.9$ and $1.3$, respectively.  Parameters: $D=3$,
    $\epsilon_0=0$ for the RLM and $E_d=-0.7$ in the SBMFT, 5000
    realizations are used, and there are 500 energy levels within the
    band.  }
\label{fig:LevelCor}
\end{figure}

To characterize the relation between the weak and strong coupling
levels, we consider the distribution of the normalized level shift
defined by 
\begin{equation}
   S \in \left\{ \frac{|\lambda_\kappa-\epsilon_i|}{|\epsilon_{i+1}-\epsilon_i|},
   \frac{|\lambda_\kappa-\epsilon_{i+1}|}{|\epsilon_{i+1}-\epsilon_i|} \right\} \;.
\end{equation}
The range of $S$ is from $0$ to $1$. 

We start by considering the non-interacting RLM, introducing the
resonant level right at the chemical potential, $\epsilon_0 = 0$, and
then analyzing those levels within the resonant width, $-\Gamma_0/2 <
\lambda_\kappa < \Gamma_0/2$.  Fig.~\ref{fig:LevelCor} shows the
probability distribution $P(S)$ obtained by sampling a large number of
realizations. We see that this distribution is independent of the
coupling strength (for levels within the resonant width). The
corresponding results for the toy model, Eqs.~(\ref{eq:PtoyGOE}) and
(\ref{eq:Ptoy2}), are plotted in Fig.~\ref{fig:LevelCor} as well. The
toy model gives a good overall picture of both the distribution of $S$
and the difference between the orthogonal and unitary cases: the
strong coupling levels are concentrated near the original levels in
the case of the GOE while they are pushed away from the original
levels in the GUE. Quantitatively, however, the weight in the middle
of the interval is greater in the full RLM than in the toy model.
Comparing the GUE case with the prediction Eq.~(\ref{eq:Ptoy2})
obtained from the second toy model (after performing the proper
averaging over $\bar \lambda/\bar \Gamma$, see
Appendix~\ref{app:averaging}), we see that this difference can be
attributed to the mean effect of the levels other than the closest
ones, which tend to push $\lambda_\kappa$ into the middle of the
interval $[\epsilon_i,\epsilon_{i+1}]$.  Remarkably, as seen in
Fig.~\ref{fig:LevelCor}, neglecting the fluctuations of the
wave-functions other than $|\phi_i(0)|^2$ and $|\phi_{i+1}(0)|^2$
tends to make this ``pressure'' toward the center somewhat bigger than
it would be if all fluctuations were taken into account.

One intriguing 
prediction of the toy model is the square root singularity at $S=0$ and $S=1$ in
the GOE case. To see whether this is present in the RLM numerical results, we
plot the cumulative distribution function on a log-log scale in the inset in
Fig.~\ref{fig:LevelCor}; the resulting straight line parallel to the toy model
result (though with slightly smaller magnitude) shows that, indeed, the square
root singularity is present. As predicted by the toy model, breaking time
reversal symmetry causes a dramatic change in $P(S)$.

Results for the full SBMFT treatment of the infinite-$U$ Anderson model are
shown in Figs.~\ref{fig:LevelCor}(c) and \ref{fig:LevelCor}(d) for
the GOE and GUE, respectively. Only levels satisfying  
$\mathcal{E}_0-\Gamma(\eta)/2 < \lambda_\kappa < \mathcal{E}_0+\Gamma(\eta)/2$ are
included; these are the levels that are within the Kondo
resonance. Fig.~\ref{fig:LevelCor} shows that the perturbed energy
levels within the Kondo resonance for the interacting model have the same
statistical properties as the ones within the resonance for the
non-interacting model.

%%%%%%%%%%%%%%%%%%%%%%%%%%%%%%%%%%%%%%%%%%%%%%%%%%%%%%%%%%%%%%%%%%%%%%%%%%%%
\section{Wave-function correlations}
\label{sec:WF}

We turn now to the properties of the eigenstates.
A key quantity of interest in quantum dot physics is the magnitude of
the wave function of a level at a point in the dot that is coupled to
an external lead. This quantity is directly related to the conductance
through the dot when the chemical potential is close to the energy of
the level \cite{Kouwenhoven97,Kaul03_kondoflucts}. We assume that the
probing lead is very weakly coupled, so the relevant quantity is the
magnitude of the wave function in 
the absence of leads. Within our RMT model, all points other than the
point $\mathbf{r}=0$, to which the impurity is coupled, are
equivalent. The evolution of the magnitude of the quasi-particle  wave
function probability 
$|\psi_{i}(\mathbf{r})|^2$, at some arbitrary point $\mathbf{r}\neq 0$
as a function of the coupling strength is shown in
Fig.~\ref{fig:EW_GOE}(c) for GOE and Fig.~\ref{fig:EW_GUE}(c) for
GUE. Note the large variation in magnitude, often over a narrow window
in coupling $V_0$, and the fact that the magnitude of each level tends
to go to 0 at some value of $V_0$ (though not all at the same value).

In order to understand how the coupling to an outside lead at  $\mathbf{r}$ is affected by the coupling to the impurity, we
study the correlation between the quasi-particle wave-function
probability $|\psi_{\kappa(i)}(\mathbf{r})|^2$ and the unperturbed
wave-function probability $|\phi_i(\mathbf{r})|^2$ 
[using the convention of Sec.\,\ref{sec:preliminary}, $\kappa(i) = i$].
More specifically, we will consider in this section the correlator
\begin{equation}
\mathcal{C}_{i,\kappa(i)} = \frac{\overline{|\phi_i(\mathbf{r})|^2
    |\psi_{\kappa(i)}(\mathbf{r})|^2}- 
\overline{|\phi_i(\mathbf{r})|^2}\cdot\overline{|\psi_{\kappa(i)}(\mathbf{r})|^2}}
   {\sigma(|\phi_i(\mathbf{r})|^2)\sigma(|\psi_{\kappa(i)}(\mathbf{r})|^2)}
   \; .
\end{equation}
The average $\overline{(\cdot)}$ here is over all realizations, for
arbitrary fixed $\mathbf{r} \neq 0$, and $\sigma(\cdot)$ is the square
root of the variance of the corresponding quantity.

\begin{figure}[t]
\centering
\includegraphics[width=3.4in,clip]{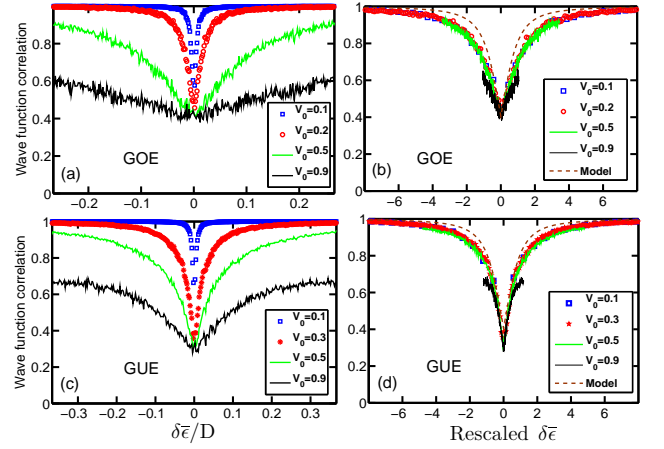}
\vspace*{-0.2in}
\caption{(Color online) Wave function correlator
  $\mathcal{C}_{i,\kappa(i)}$ for the non-interacting RLM 
  ($\bar\eta=1$).
  (a) GOE and (c) GUE, as a function of the average distance between
  $\epsilon_i$ and $\epsilon_0$. (b) GOE and (d) GUE, as a function of
  rescaled average distance $\epsilon=(i \Delta- D/2)/\Gamma_0$.
  The dashed lines are the result of Eq.(\ref{eq:Cikappa})  in which
    the wavefunction fluctuations are taken into account but the
    energy levels are assumed bulk-like.  Parameters: $D=3$, impurity
  energy level $\epsilon_0 =0$, $5000$ realizations, and $500$ energy
  levels within the band.  
}
\label{fig:NIAM_WF}
\end{figure}

We expect that, as for the energies, most of the wave-function
fluctuation properties can be understood by starting from the RLM
Eq.~(\ref{eq:RLM}) despite the fact that fluctuations of the
mean-field parametersmare not included.  We start therefore with
Fig.~\ref{fig:NIAM_WF} which shows $\mathcal{C}_{i,\kappa(i)}$ for the
non-interacting RLM as a function of the average distance $\delta \bar
\epsilon_i = (i \Delta- D/2)$ between $\epsilon_i$ and $\epsilon_0 =0$
(which is in the middle of the band). In Fig.~\ref{fig:NIAM_WF} (a)
and (c), the correlator $\mathcal{C}_{i,\kappa(i)}$ has a dip at the
position of the impurity level.  The width of the dip increases as the
coupling $V_0$ increases. Rescaling the energy axis by $\Gamma_0$, as
done in Fig.~\ref{fig:NIAM_WF} (b) and (d), shows that the width of
the dip is proportional to the resonance width. One also finds that
$\mathcal{C}_{i,\kappa(i)}$ is $\simeq 1$ for the energy levels
outside the resonance, which is expected, but that
$\mathcal{C}_{i,\kappa(i)}$ is slightly below 1/2 in the center of the
resonance.

\begin{figure}[t]
\centering
\vspace*{0.2in}
\includegraphics[width=3.4in,clip]{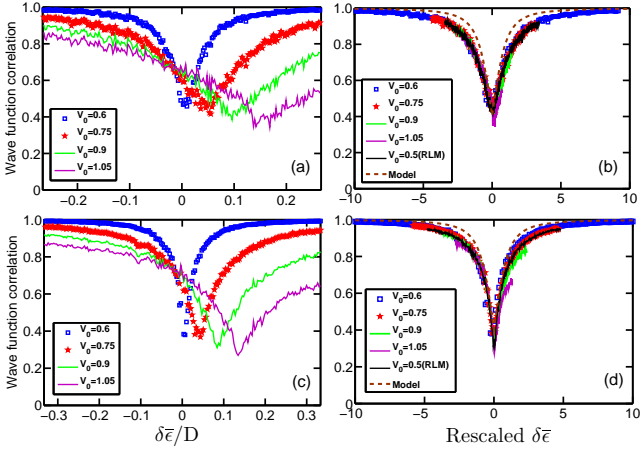}
\vspace*{-0.2in}
\caption{(Color online) Wave function correlation,
  $\mathcal{C}_{i,\kappa(i)}$, for the SBFMT approach to the
  infinite-$U$ Anderson model. 
  (a) GOE and (c) GUE, as a function of the average
  distance from the middle of the band. (b) GOE and (d) GUE,
  as a function of rescaled average distance $\delta \bar
  \epsilon=[(i\Delta \!- \!D/2) - (\mathcal{E}_0(\xi)\!
  -\!\mu)]/\Gamma(\eta)$.  The black lines labeled RLM are results for
  the non-interacting RLM at $V_0=0.5$; 
 The dashed lines are the result of Eq.(\ref{eq:Cikappa})  in which
    the wavefunction fluctuations are taken into account but the
    energy levels are assumed bulk-like.
  Parameters: $D=3$, impurity energy level
  $E_d=-0.7$, $5000$ realizations, and $500$
  energy levels within the band.
  }
\label{fig:SBMFT_WF}
\end{figure}

Turning now to the full self-consistent problem, we plot in
Fig.~\ref{fig:SBMFT_WF} the wave-function correlator
$\mathcal{C}_{i,\kappa(i)}$ for the full SBMFT approach to the
infinite-$U$ Anderson model. Panels (a) and (c) show that the
wave-function correlation has a dip similar to that in the RLM
results. The dip is located at $\delta \bar \epsilon_i=0.0$ for small
coupling (i.e.\ $V_0=0.6$), and then moves to larger $\delta \bar
\epsilon_i$ as the coupling $V_0$ increases.  This is a natural result
for the highly asymmetric infinite-$U$ Anderson model: for small
coupling, the SBMFT calculation leads to $\mathcal{E}_0 -\mu = E_d-\xi
\approx 0$, while for increasing $V_0$, $\mathcal{E}_0 -\mu$ increases
to positive values. In fact, the dip corresponds to the effective
Kondo resonance.  Incorporating the shift of $\mathcal{E}_0(\xi)$ and
rescaling by $\Gamma(\eta)\sim T_K$, we plot the wave function
correlation as a function of $\delta \tilde \epsilon \equiv [(i\Delta
\!- \!D/2) - (\mathcal{E}_0(\xi)\!  -\!\mu)]/\Gamma(\eta)$ in
Fig.~\ref{fig:SBMFT_WF} (b) and (d).  All the curves collapse onto
universal curves, one for the GOE and another for the GUE. In
addition, the universal curves are the same as the universal curves
for the RLM.

As anticipated, the (fixed parameter) resonant level model contains
essentially all the physics controlling the behavior of the correlator
$\mathcal{C}_{i,\kappa(i)}$.  We can therefore try to understand the
behavior of this quantity without taking into account the fluctuations
of the mean field parameters.

Using again the Green function Eq.~(\ref{eq:Green}), we can define the
quasi-particle wave-function probability $|\psi_\kappa({\bf r})|^2$ as
the residue at $\lambda_\kappa$ of $\langle {\bf r} | \hat G | {\bf r}
\rangle = \sum_{jj'} \psi_j({\bf r}) G_{jj'} \psi_{j'}({\bf r})$.
From the expression for $G_{jj'}$ given in
Eq.~(\ref{AndersonboxGll1}), we thus have 
\begin{equation}
\label{eq:wv}
|\psi_\kappa({\bf r})|^2 =  \sum_{jj'} \frac{\psi_j({\bf r}) v^*_j
}{\lambda_\kappa - \epsilon_j}  \cdot u_\kappa \cdot 
\frac{\psi^*_{j'}({\bf r}) v_{j'}}{ \lambda_\kappa - \epsilon_{j'}} \; ,
\end{equation}
where $v_j = \eta V_0 \phi_j({\bf 0})$ is the coupling of the state
$j$ to the impurity and $u_\kappa \equiv |\langle
\psi_\kappa|f\rangle|^2$ is given by Eq.~(\ref{eq:uk}).  Therefore 
\begin{equation}
|\psi_\kappa({\bf r})|^2 \cdot |\phi_i({\bf r})|^2 = \sum_{jj'}
\Omega^\kappa_{jj'} \, \psi_j({\bf r}) \psi^*_{j'}({\bf r})\psi_i({\bf r})
\psi^*_{i}({\bf r}) \; , 
\end{equation}
where we have defined
\begin{equation}
\Omega^\kappa_{jj'} \equiv \frac{ v^*_j}{\lambda_\kappa - \epsilon_j}  
\cdot u_\kappa \cdot \frac{v_{j'}}{\lambda_\kappa - \epsilon_{j'}}  \; .
\end{equation}

In our random matrix model, there is no correlation between different
wave-functions or between wave-functions and energy levels.  We thus
have
\begin{equation}
\overline{|\phi_i(\mathbf{r})|^2 |\psi_{\kappa}(\mathbf{r})|^2}- 
\overline{|\phi_i(\mathbf{r})|^2} \cdot
\overline{|\psi_{\kappa}(\mathbf{r})|^2}  
=
\sum_{jj'} \overline{\Omega^\kappa_{jj'}} \cdot g_{iijj'}
\end{equation}
where
\begin{eqnarray}
g_{ii'jj'} & \equiv & \left[\overline{\psi_i({\bf r})
\psi^*_{i'}({\bf r}) \psi_j({\bf r}) \psi^*_{j'}({\bf r})} \right. \nonumber \\
& & - \left.
\overline{\psi_i({\bf r}) \psi^*_{i'}({\bf r})} \cdot
 \overline{\psi_j({\bf r}) \psi^*_{j'}({\bf r})} \right]\; .
\end{eqnarray}
Because the wave-functions are independent and Gaussian
distributed,  $g_{ii'jj'} = (2/\beta) \delta_{ii'}\delta_{jj'}\delta_{ij} 
\overline{|\psi_i({\bf r})|^2}^2$ 
(remembering the normalization $\overline{|\psi_i({\bf r})|^2} = 1/N$ and $\beta=1$ for GOE while $\beta=2$ for GUE).  In the same way, we have 
$\sigma(|\phi_i(\mathbf{r})|^2) = {g_{iiii}} = (2/N\beta)$.
Furthermore, using Eq.~(\ref{eq:wv}) and the limit
$\Gamma \gg \Delta$, we have
$\sigma(|\psi_{\kappa}(\mathbf{r})|^2) \simeq (2/N\beta)$ which then yields
\begin{equation} \label{CikFromOmega}
\mathcal{C}_{i,\kappa} = \overline{\Omega^\kappa_{ii}}
= \overline{u_\kappa \cdot \frac{|v_i|^2}{(\lambda_\kappa - \epsilon_i)^2}}
  \; .
\end{equation}
[As a side remark, we note that by differentiating Eq.~(\ref{eq:lambdas}) with
respect to $\epsilon_i$, one can show that $\partial \lambda^2_\kappa
/ \partial \epsilon_i = \Omega^\kappa_{ii}$, and thus
$\mathcal{C}_{i,\kappa} =\overline{\partial \lambda_\kappa
/ \partial \epsilon_i}$.]

A good approximation to $\mathcal{C}_{i,\kappa(i)}$ can then be
obtained from the bulk-value, using Eq.~(\ref{ukappabulk}) to evaluate
(\ref{CikFromOmega}) in the bulk limit yields
\begin{equation}
\left(\Omega^\kappa_{ii}\right)^{\rm bulk} \equiv
\Big[\displaystyle \delta_\kappa^2 \sum_i \frac{1}{(i+\delta_\kappa)^2}\Big]^{-1} \; ,
\end{equation}
where $\delta_\kappa = (\lambda_{\kappa(i)} - \epsilon_i)/\Delta$.
Using Eqs.~(\ref{app:deltak}), (\ref{app:sum1}), and (\ref{app:sum2})
from Appendix A, we thus have
\begin{equation} \label{eq:Cikappa}
\mathcal{C}_{i,\kappa(i)} \simeq \frac{1}{\left[{\rm
      cotan}^{-1}\left({\delta \bar \epsilon_i}/{\Gamma} \right)
  \right]^2 \left(1 + \left( {\delta \bar \epsilon_i}/{\Gamma}
    \right)^2 \right)}
  \; ,
\end{equation}
which, as anticipated, depends only on the ratio $({\delta \bar
  \epsilon_i}/{\Gamma})$.  The curve resulting from this expression is
shown in Figs.~\ref{fig:NIAM_WF} and \ref{fig:SBMFT_WF} and is in good
agreement with the numerical data.

Eq.~(\ref{eq:Cikappa}) provides a good qualitative and quantitative
description of the energy dependence of the correlator
$\mathcal{C}_{i,\kappa(i)}$ [although differences between 
$\left(\Omega^\kappa_{ii}\right)^{\rm bulk}$ and 
$\overline{\Omega^\kappa_{ii}}$ are visible].  
In a conductance experiment, however, only
the levels near the Fermi energy that are within the Kondo resonance
contribute to the conductance. In the middle of the resonance,
$\mathcal{C}_{i,\kappa(i)}$ is slightly less than one half.  
\textit{At
temperatures lower than the mean spacing $\Delta$, for which only one
state would contribute to the conductance, there would be some
correlation, but only a partial one, between the fluctuations of the
conductance in the uncoupled system and the one in the Kondo limit.}

%%%%%%%%%%%%%%%%%%%%%%%%%%%%%%%%%%%%%%%%%%%%%%%%%%%%%%%%%%%%%%%%%%%%%%%%%%%%
\section{Discussion and Conclusions}
\label{sec:conclusion}

We have obtained results for the correlation between the statistical
fluctuations of the properties of the reservoir-dot electrons in two
limits: the high-temperature non-interacting gas on the one hand
($T\gg \Tk^{\rm bulk}$) and, on the other hand, the quasiparticle gas
when the Anderson impurity is strongly coupled ($T \ll \Tk^{\rm
  bulk}$).  The exact treatment of the mesoscopic Kondo problem in the
low temperature regime is, however, nontrivial. Since the very low
temperature regime ($T\ll \Tk^{\rm bulk}$) is described by a
Nozi\`eres-Landau Fermi liquid, we tackled this problem by using the
slave boson mean field approximation, through which the infinite-$U$
Anderson model is mapped to an effective resonant level model with
renormalized impurity energy level and coupling.

We derived the spectral joint distribution function, 
Eq.~(\ref{eq:joint_final}), which
in principle contains all the information about the
correlations between the high and low temperature spectra of the
mesoscopic Anderson box. In the spirit of the Wigner surmise, a
solvable toy model was introduced to avoid the complications of the
joint distribution function. The toy model provides considerable insight into
the spectral correlations in the original model. 

The numerical infinite-$U$ SBMFT calculation shows the following
results.  First, the distributions of the mean field parameters are
Gaussian. Second, the distribution of the critical coupling $J_K^c$
does not vanish at zero which shows that there exist some realizations
for which the Kondo effect appears at any bare coupling $V_0$ and
impurity energy level $E_d$. Third, for the GOE, the spectral spacing
distribution has two sharp peaks at $S=0$ and $S=1$, showing that the
two perturbed energy levels (i.e.\ those for $T\ll \Tk^{\rm bulk}$)
are close to the unperturbed ones ($T\gg \Tk^{\rm bulk}$).  For the
GUE, the peak of the spectral correlation function is located at
$S=0.5$ corresponding to the center of the two unperturbed energy
levels. In addition, the spectral spacing distribution for different
coupling strengths $V_0$ collapse to universal forms, one for GOE and
one for GUE, when we consider only energy levels within the Kondo
resonance.

Finally, we studied
the influence of the Anderson impurity on the coupling strength between an
outside lead and the energy levels of the large dot, as would be probed in a conductance measurement. This is
characterized by the intensity of the wave function at an arbitrary point. The
correlation function of this intensity corresponding
to the unperturbed system and perturbed system shows a dip located at
the Kondo resonance, and the width of the dip is proportional to the
width of the Kondo resonance. Only the part of the wave
function amplitude that corresponds to the perturbed energy levels
within the Kondo resonance will be significantly affected due to the
coupling to the Kondo impurity.

%%%%%%%%%%%%%%%%%%%%%%%%%%%%%%%%%%%%%%%%%%%%%%%%%%%%%%%%%%%%%%%%%%%%%%%%%%%%
\section*{Acknowledgments}

The work at Duke was supported by U.S.\,DOE, Office of Basic Energy
Sciences, Division of Materials Sciences and Engineering under Grant
No.~\#DE-SC0005237 (D.E.L. and H.U.B.).

\appendix

\section{Kondo temperature and values of the mean field 
parameters in  the bulk limit}
\label{app:bulk}

In this appendix, we provide a brief reminder of the derivation of the
Kondo temperature and mean field parameters in the bulk limit.  We
define this latter by taking $N \to \infty$ and assuming that there
are no fluctuations in either the wave-functions nor the unperturbed
levels: for all $i$, $x_i =1$ and $\epsilon_{i+1} - \epsilon_{i} =
\Delta$.  We further assume the chemical potential $\mu$ in the middle
of the band.

Under these assumptions, the equation
defining the Kondo temperature, (\ref{eq:TKAndersonbox2}), reads
\begin{equation}
\frac{E_{d}}{\rho_0 V_0^{2}}
= \int_{-D/2}^{+D/2} 
\frac{dy}{y} \tanh{[y/2T_{K}]} = 2 \ln\left(\frac{a_K}{2}
  \frac{D}{T_K} \right) 
\; ,
\end{equation}
%=======
%just for info : 
%  \ln(a_K/2} == [ \int_0^1 (dy/y) tanh(y) + \int_1^\infty (dy/y)
%  (tanh(y)-1) == 0.81878
%  ==> a_K = 2.2677.. / 2
%
%=======
($a_K \simeq 1.1338..$), and thus
\begin{equation}
T_K = \frac{a_K}{2} D \exp\left(-\frac{|E_d|}{2\rho_0 V_0^2} \right) \;.
\end{equation}

Turning now to the (zero-temperature) mean-field parameters, we shall
denote their value in the bulk limit by $\bar \eta$ and $\bar \xi$, and by
$\bar \Gamma \equiv \Gamma(\bar \eta, \bar \xi)$ and $ \mathcal{\bar E}_0
\equiv \mathcal{E}_0(\bar \eta, \bar \xi)$ the corresponding width and
center of the resonance.  Let us consider the perturbed eigenlevel
$\lambda_\kappa \in [\epsilon_i,\epsilon_{i+1}]$ , and $\delta_\kappa
\equiv (\lambda_\kappa \! - \! \epsilon_i)/\Delta$.
Eq.~(\ref{eq:lambdas}) reads in the bulk limit
\begin{equation} \label{app:deltak}
\frac{\lambda_\kappa - \mathcal{\bar E}_0}{\bar \Gamma} 
= \frac{1}{\pi}  \sum_{j}\frac{1}{\delta_\kappa-j}  \; ,
\end{equation}
and likewise Eq.~(\ref{eq:uk}) for the overlap 
$u_\kappa = |\langle \psi_\kappa | f \rangle |^2$  is 
(assuming $\bar \Gamma \gg \Delta$)
\begin{equation} \label{ukappabulk}
u_\kappa = 
  \frac{\pi \Delta}{\bar \Gamma}  \frac{1}{\sum_{j} {(\delta_\kappa -
      j)^{-2}}} \;. 
\end{equation}

Using the identities
\begin{eqnarray} 
   \sum_{j}  \frac{1}{\delta_\kappa-j} &=& 
         \pi\, {\rm cotan}(\pi \delta_\kappa) \; , \label{app:sum1} \\
   \sum_{j} \frac{1}{ (\delta_\kappa -j)^{2}}  & = & 
        \pi^2 [1 + {\rm cotan}^2(\pi \delta_\kappa)] \; , \label{app:sum2} 
\end{eqnarray}
together with Eq.~(\ref{app:deltak}), one obtains
\begin{equation}
 \sum_{j} \frac{1}{ (\delta_\kappa
   -j)^{2}}=\pi^2[1+\frac{(\lambda_\kappa - \mathcal{\bar
     E}_0)^2}{\bar \Gamma^2} ] \;. 
\end{equation}
We therefore can express the bulk analogues $I^{\rm bulk}(\bar
\eta,\bar \xi)$ and $J^{\rm bulk}(\bar \eta,\bar \xi)$ of the sums
introduced in the mean-field equations
(\ref{Consistenteta6})-(\ref{Consistentxi6}) as
\begin{eqnarray}
I^{\rm bulk}(\bar \eta,\bar \xi) & \equiv & 
\int_{-D/2}^{+D/2} dy 
\frac{{\rm sgn}(y \!- \! \delta \bar \xi )\,y}{y^2 + \bar
  \Gamma^2} \nonumber \\ 
 & = & 2 \ln \left( \frac{1}{\sqrt{1 +
      (\delta \bar \xi/\bar \Gamma)^2}} \frac{D}{2 \bar \Gamma} \right)
\label{app:Ibulk} \\
J^{\rm bulk}(\bar \eta,\bar \xi) &\equiv& \frac{1}{\pi \bar \Gamma} 
\int_{-\infty}^{\delta \bar \xi}   \frac{dy}{ 1 + (y/\bar \Gamma)^2}
\nonumber \\ 
 & = & \frac{1}{2} + \frac{1}{\pi}
\tan^{-1}({\delta \bar \xi}/{\bar \Gamma}) \; .  \label{app:Jbulk}
\end{eqnarray}
with $\delta \bar \xi \equiv (\bar \xi\! - \! E_d) = (\mathcal{\bar
  E}_0\! - \!  \mu)$.

Eq.~(\ref{app:Jbulk}) inserted into (\ref{Consistentxi6}) yields
\begin{equation} \label{app:deltabarxi}
\delta \bar \xi/\bar \Gamma = - \tan(\pi\eta^2/2) \; ,
\end{equation}
which in the  Kondo regime $(\eta \ll 1)$ implies $\delta \bar
\xi/\bar \Gamma = O(\eta^2)$.  Inserting Eq.~(\ref{app:Ibulk}) into
(\ref{Consistenteta6}) then gives
\begin{equation} \label{app:barGamma}
\bar \Gamma = \frac{D}{2} \exp \left( - \frac{|E_d|}{2\rho_0 V_0^2}
\right)
\end{equation}
Thus  in this regime  $\Tk$ and $\bar \Gamma$ differ just by the factor $a_K
\simeq 1.133$.  In the mixed valence regime 
 $T_K/\bar \Gamma  = a_K \sqrt{1 + \tan^2(\pi\eta^2/2)}$, which however
 remains of order one as long as $(1\!-\!\eta^2)$ does.

As a final comment, we note that Eq.~(\ref{app:barGamma}) implies
$\eta^2 = (D^2/2\pi V_0^2) \exp(- |E_d|D/2V_0^2)$, from which we
obtain an explicit condition
\begin{equation} \label{eq:barGamma}
\exp \left( - \frac{1}{\rho_0 J_K}\right) \ll 2\pi
\frac{V_0^2}{D^2}
\end{equation}
to be in the  Kondo regime.

\section{Constraint on the sum of the eigenvalues of the resonant
  level model}
\label{app:EVconstraint}

In this appendix, we briefly demonstrate
Eq.~(\ref{eq:EL_sum_condition}) constraining the sum of the
eigenvalues of the RLM.

Starting from $\langle \psi_\kappa|H_{\rm
  RLM}|\psi_\kappa\rangle=\lambda_\kappa\langle
\psi_\kappa|\psi_\kappa\rangle$ we may insert the identity
$I=\sum_{i=0}^N |i\rangle\langle i|$ on the right hand side (with the
notation that $|i=0\rangle \equiv |f\rangle$) and obtain
\begin{eqnarray}
 \lefteqn{\sum_{i=0}^N(\lambda_\kappa-\epsilon_i) \langle
   \psi_\kappa|i\rangle \langle i|\psi_\kappa\rangle} && \\\nonumber 
 &&= v\bigg( \sum_{i=0}^N \phi_i(0)  \langle f|\psi_\kappa\rangle
 \langle \psi_\kappa|i\rangle + {\rm h.c.}\bigg) \;. 
\end{eqnarray}
The sum of these equations, $\sum_{\kappa=0}^N$, is
\begin{eqnarray}
\label{eq:EL_sum_app}
\sum_{\kappa=0}^N \lambda_\kappa - \sum_{i=0}^N \epsilon_i &=& v\Big(
\displaystyle\sum_{i,\kappa=0}^N \phi_i(0)  \langle f|\psi_\kappa\rangle \langle 
\psi_\kappa|i\rangle + {\rm
  h.c.}\Big)\nonumber\\ 
&=& v\Big( \displaystyle\sum_{i=0}^N \phi_i(0)\delta_{0i} + {\rm
  h.c.}\Big)\nonumber\\ 
&=& 0 \;;
\end{eqnarray}
thus, the sum of the two sets of eigenvalues must be equal.

\section{Averaging of Eq.(\ref{eq:Ptoy2})}
\label{app:averaging}

Averaging Eq.~(\ref{eq:Ptoy2}) over the variable
$\Lambda \equiv \bar \lambda/\Gamma$ in some range  $[0,\Lambda_{\rm
  max}]$, we find after a bit of algebra
\begin{eqnarray*}
\frac{1}{\Lambda_{\rm  max}} && \int_0^{\Lambda_{\rm  max}} P(\sigma)
= 
\frac{4/\pi}{1+2\sigma} \exp \left[-\tilde f(\sigma) \right] \times \\
&& \Big[\left(1 + (1-4\sigma^2) \tilde f'(\sigma)
+\frac{1-2\sigma}{1+2\sigma}\right) \sinh[\pi
(\sigma+\frac{1}{2})\Lambda_{\rm max}] \\ 
&& - \; \; (\sigma-\frac{1}{2}) \tilde f(\sigma)\sinh[\pi
(\sigma+\frac{1}{2}) \Lambda_{\rm 
  max}] \\ 
&& + \, \pi  (\sigma-\frac{1}{2}) \Lambda_{\rm   max}\cosh[\pi
(\sigma+\frac{1}{2}) \Lambda_{\rm 
  max}] \Big] \; , 
\end{eqnarray*}
with $\tilde f(\sigma) =  f(\sigma,\Lambda\!=\!0)$.

\bibliography{kondo,rmt,nano}

\end{document}